\documentclass[10pt,epsfig]{article}

\usepackage{multicol}
\usepackage{graphicx}
\usepackage{booktabs}
\usepackage{amssymb,bm,mathrsfs,bbm,amscd}
\usepackage[tbtags]{amsmath}
\usepackage{lastpage}
\usepackage{lscape}
\usepackage[rotateright]{rotating}

\begin{document}
\title{Forward-backward multiplicity correlations of target fragments in nucleus-emulsion collisions at a few hundred MeV/nucleon
\thanks{Submitted to Chin. Phys. C}}

\author{Dong-Hai Zhang\thanks{Corresponding author. Tel: +863572051347; fax: +863572051347. E-mail address:zhangdh@dns.sxnu.edu.cn}, Yan-Ling Chen, Guo-Rong Wang, Wang-Dong Li, Qin Wang\\
Ji-Jie Yao, Jian-Guo Zhou, Rong Li, Jun-Sheng Li, Hui-Ling Li\\
Institute of Modern Physics, Shanxi Normal University, Linfen 041004, China}

\date{}
\maketitle

\begin{center}
\begin{minipage}{140mm}
\vskip 0.4in
\begin{center}{\bf Abstract}\end{center}
{The forward-backward multiplicity and correlations of target evaporated fragment(black track particle) and target recoiled proton(grey track particle) emitted in 150 A MeV $^{4}$He-emulsion, 290 A MeV $^{12}$C-emulsion, 400 A MeV $^{12}$C-emulsion, 400 A MeV $^{20}$Ne-emulsion and 500 A MeV $^{56}$Fe-emulsion interactions are investigated. It is found that the forward and backward averaged multiplicity of grey, black and heavily ionized track particle increase with the increase of target size. Averaged multiplicity of forward black track particle, backward black track particle, and backward grey track particle do not depend on the projectile size and energy, but the averaged multiplicity of forward grey track particle increases with increase of projectile size and energy. The backward grey track particle multiplicity distribution follows an exponential decay law and the decay constant decreases with increase of target size. The backward-forward multiplicity correlations follow linear law which is independent of the projectile size and energy, and the saturation effect is observed in some heavy target data sets.}\\

\end{minipage}
\end{center}

\vskip 0.4in
\baselineskip 0.2in
\section{Introduction}

The study of the production of backward particles in hadron-nucleus and nucleus-nucleus interactions at high energies has received considerable experimental and theoretical attention because the fact that the backward emission of relativistic particles in high energy hadron-nucleon collisions is kinematically restricted. The backward emission of protons and pions in hadron-nucleus[1-7] and nucleus-nucleus[8-18] at relativistic energies have been investigated exclusively. It is found that the backward proton production was attributed to the absorption of secondary pions by a nucleon pair in the target nucleus, and the backward pions production shown to be consistent with the cumulative effect\cite{bald}. The emission of target evaporated fragments is isotropic in the rest system of target nucleus according to the cascade evaporation model\cite{powe}, but attributed to the electromagnetic field from projectile the emission of target evaporated fragments is close to $\theta\approx90^{\circ}$ symmetric and the multiplicity in forward hemisphere is greater than that in backward hemisphere. There is no reasonable theoretical model to explain the salient features of forward-backward multiplicity correlations. For the particle production in backward hemisphere and forward-backward multiplicity correlations in nucleus-nucleus collisions at intermediate and high energies (a few hundreds MeV/nucleon), a little attention is paid to study.

According to the participant-spectator model\cite{bowm} of the high energy nucleus-nucleus collisions, the projectile and target sweep out cylindrical cuts through each other. The overlapping region of nuclear volumes is called the participant region, where multiple production of new particles occurs and the nuclear matter breaks up into nucleons. The remaining parts of nuclei which do not participate in the disintegration process are called the spectator regions of the projectile and target nuclei. In a central collision the projectile is drilling a cylindrical hole through the target nucleus, striking every nucleon in its path. Some of the struck nucleons will penetrate into the spectator part whereas some will escape through the hole. It is assumed that effectively only those nucleons that are originating from the surface region of the cylinder penetrates into the spectator and that these nucleons move away from the centre of the hole. In a peripheral or semi-central collision only a part of the cylindrical hole is developed, and here the probability that a struck target nucleon will disappear without penetrating into the spectator increases with the decrease of the collision centrality. During this colliding process, a fraction of the available energy is transferred to the spectator parts of colliding nuclei, leaving those nuclear remnants in an excited state. After this stage, the de-excitation of the nuclear remnants take place and the target and projectile fragments are formed. In general, this reaction mechanism is also reasonable for the intermediate and high energy, but the production of new particles in participant region is highly suppressed because of the limited reaction energy.

In this paper the forward-backward multiplicity and correlations of target evaporated fragment and target recoiled proton produced in 150 A MeV $^{4}$He-emulsion, 290 A MeV $^{12}$C-emulsion, 400 A MeV $^{12}$C-emulsion, 400 A MeV $^{20}$Ne-emulsion and 500 A MeV $^{56}$Fe-emulsion interactions are investigated, we want to find out the general characteristics of the particle production in backward hemisphere and forward-backward multiplicity correlations in nucleus-nucleus collisions at intermediate and high energies.

\section{Experimental details}

Five stacks of nuclear emulsion made by Institute of Modern Physics, Shanxi Normal University, China, are used in present investigation. The emulsion stacks were exposed horizontally at HIMAC, NIRS, Japan. The beams were 150 A MeV $^{4}$He, 290 A MeV $^{12}$C, 400 A MeV $^{12}$C, 400 A MeV $^{20}$Ne and 500 A MeV $^{56}$Fe respectively, and the flux was 3000 ions/$cm^{2}$. BA2000 and XSJ-2 microscopes with a 100$\times$ oil immersion objective and 10$\times$ ocular lenses were used to scan the plates. The tracks were picked up at a distance of 5 mm from the edge of the plates and were carefully followed until they either interacted with emulsion nuclei or escaped from the plates. Interactions which were within 30 ${\mu}m$ from the top or bottom surface of the emulsion plates were not considered for final analysis. All the primary tracks were followed back to ensure that the events chosen do not include interactions from the secondary tracks of other interactions. When they were observed to do so the corresponding events were removed from the sample.

In each interaction all of the secondaries were recorded which include shower particle, target recoiled proton, target evaporated fragment and projectile fragments. According to the emulsion terminology\cite{powe}, the particles emitted from high energy nucleus-emulsion interactions are classified as follows.

(a) Black track particle ($N_{b}$). They are target evaporated fragments with ionization $I>9I_{0}$, $I_{0}$ being the minimum ionization of a single charged particles. Range of black particle in nuclear emulsion is $R<3$ mm, velocity is $v<0.3c$, and energy is $E<26$ MeV. The multiplicity of black track particle emitted in forward (emission angle $\theta\leq90^{\circ}$)and backward hemisphere ($\theta>90^{\circ}$) is denoted as $n_{b}^{f}$ and $n_{b}^{b}$ respectively, and total multiplicity is denoted as $n_{b}$.

(b) Grey track particle ($N_{g}$). They are mostly recoil protons in the kinetic energy range $26\leq{E}\leq375$ MeV and a few kaons of kinetic energies $20\leq{E}\leq198$ MeV and pions with kinetic energies $12\leq{E}\leq56$ MeV. They have ionization $1.4I_{0}\leq{I}\le9I_{0}$. Their ranges in emulsion are greater than 3 mm and have velocities within $0.3c\leq{v}\leq0.7c$. The multiplicity of grey track particle emitted in forward and backward hemisphere is denoted as $n_{g}^{f}$ and $n_{g}^{b}$ respectively, and total multiplicity is denoted as $n_{g}$.

The grey and black track particles together are called heavy ionizing particles ($N_{h}$). The multiplicity of heavy ionizing particle emitted in forward and backward hemisphere is denoted as $n_{h}^{f}$ and $n_{h}^{b}$ respectively, and total multiplicity is denoted as $n_{h}$.

(c) Shower particle ($N_{s}$). They are produced single-charged relativistic particles having velocity $v\geq{0.7c}$. Most of them belong to pions contaminated with small proportions of fast protons and K mesons. It should be mentioned that for nucleus-emulsion interactions at a few hundred MeV/nucleon most of shower particles are projectile protons not pions.

(d) The projectile fragments ($N_{f}$) are a different class of tracks with constant ionization, long range, and small emission angle.

The nuclear emulsion is composed of a homogeneous mixture of nuclei. The chemical composition of nuclear emulsion is H, C, N, O, S, I, Br, and Ag, and major composition is H, C, N, O, Br, and Ag. According to the value of $n_{h}$ the interactions are divided into following three groups.

Events with $n_{h}\leq1$ are due to interactions with H target and peripheral interactions with CNO and AgBr targets.

Events with $2\leq{n_{h}}\leq7$ are due to interactions with CNO targets and peripheral interactions with AgBr targets.

Events with $n_{h}\geq8$ definitely belong to interactions with AgBr targets.

\section{Results and discussion}

Fig. 1 shows the multiplicity distribution of backward grey track particles in different type of  150 A MeV $^{4}$He, 290 A MeV $^{12}$C, 400 A MeV $^{12}$C, 400 A MeV $^{20}$Ne and 500 A MeV $^{56}$Fe induced nuclear emulsion interactions. It is found that the probability decreases with the increase of $n_{g}^{b}$, which can be well represented by a exponential decay law with the form of $p(n_{g}^{b})=a\exp({-\lambda{n_{g}}^{b}})$.  The fitting parameters and event statistics ($N_{events}$)are presented in table 1. The decay constant $\lambda$ decreases with increase of target size, and no evident projectile size and energy is found in present investigation.

\begin{table}
\begin{center}
Table 1.The exponential fitting parameters of $n_{g}^{b}$ distribution in different type of nucleus-emulsion interactions.\\
\begin{tabular}{lccc}\hline
Type of interaction  &      a          & $\lambda$       &  $N_{events}$  \\\hline
150 A MeV $^{4}$He-CNO         & $0.761\pm0.051$ & $1.428\pm0.094$ &   272 \\
150 A MeV $^{4}$He-Em          & $0.737\pm0.044$ & $1.348\pm0.081$ &   360 \\
150 A MeV $^{4}$He-AgBr        & $0.521\pm0.094$ & $0.781\pm0.147$ &   50 \\
290 A MeV $^{12}$C-CNO         & $0.799\pm0.026$ & $1.689\pm0.045$ &   1086 \\
290 A MeV $^{12}$C-Em          & $0.715\pm0.019$ & $1.264\pm0.035$ &   432 \\
290 A MeV $^{12}$C-AgBr        & $0.466\pm0.027$ & $0.660\pm0.032$ &   1850 \\
400 A MeV $^{12}$C-CNO         & $0.911\pm0.045$ & $2.583\pm0.136$ &   450 \\
400 A MeV $^{12}$C-Em          & $0.802\pm0.032$ & $1.667\pm0.082$ &   250 \\
400 A MeV $^{12}$C-AgBr        & $0.601\pm0.045$ & $0.921\pm0.064$ &   799 \\
400 A MeV $^{20}$Ne-CNO        & $0.735\pm0.031$ & $1.365\pm0.048$ &   676 \\
400 A MeV $^{20}$Ne-Em         & $0.660\pm0.001$ & $1.080\pm0.033$ &   277 \\
400 A MeV $^{20}$Ne-AgBr       & $0.447\pm0.032$ & $0.637\pm0.038$ &   1093 \\
500 A MeV $^{56}$Fe-CNO        & $0.880\pm0.039$ & $2.221\pm0.102$ &   540 \\
500 A MeV $^{56}$Fe-Em         & $0.728\pm0.024$ & $1.329\pm0.048$ &   558 \\
500 A MeV $^{56}$Fe-AgBr       & $0.561\pm0.028$ & $0.826\pm0.035$ &   1321 \\\hline
\end{tabular}
\end{center}
\end{table}

\begin{figure}[htbp]
\begin{center}
\includegraphics[width=1.0\linewidth]{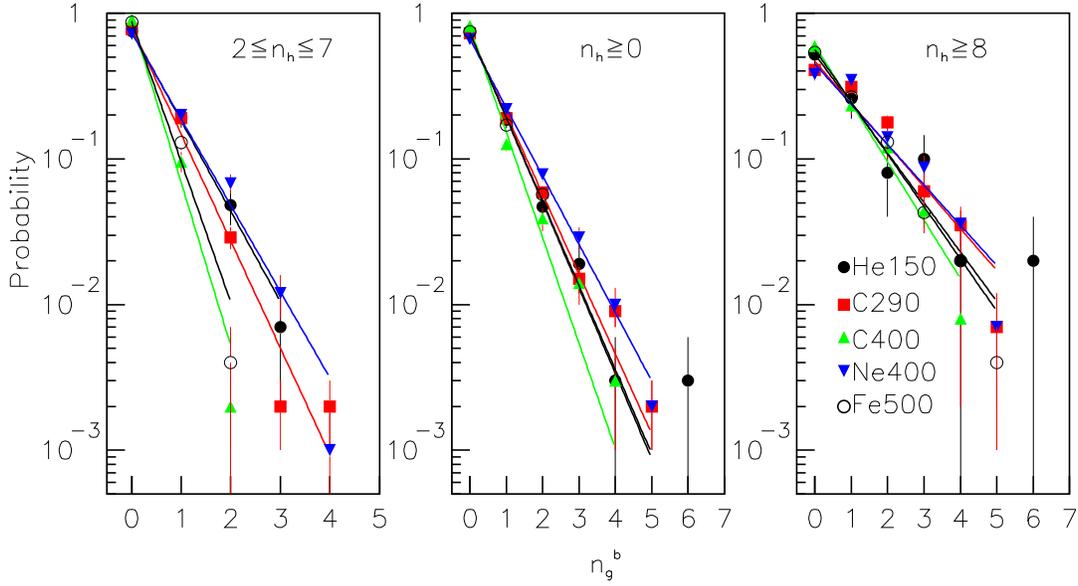}
\caption{(Color online) Multiplicity distributions of backward grey track particle in different type of interactions, the smooth curve is fitted by a exponential distribution.}
\end{center}
\end{figure}

Table 2 presents the averaged forward and backward multiplicity of black, grey, and heavy ionizing track particles in different type of 150 A MeV $^{4}$He-emulsion, 290 A MeV $^{12}$C-emulsion, 400 A MeV $^{12}$C-emulsion, 400 A MeV $^{20}$Ne-emulsion and 500 A MeV $^{56}$Fe-emulsion interactions. It is found that the value of $<n_{b}^{f}>$, $<n_{b}^{b}>$, and $<n_{g}^{b}>$ increases with the target size, and which is independent of the projectile size and energy. The value of $<n_{g}^{f}>$ is also increased with the target size and independent of the projectile size and energy except for the case of 500 A MeV $^{56}$Fe-emulsion interactions. The ratio of forward-backward multiplicity for grey track particle ($(F/B)_{g}$)decrease with the increase of target size, but for black track particle it is independent of the target size. The forward-backward multiplicity ratio for grey track particle is greater than that for black track particle, which is reasonable consistence with the prediction of the cascade evaporation model.

Fig. 2 shows the correlation between $<n_{g}^{b}>$, $<n_{g}^{f}>$ and $n_{b}$ for different type of 150 A MeV $^{4}$He-emulsion, 290 A MeV $^{12}$C-emulsion, 400 A MeV $^{12}$C-emulsion, 400 A MeV $^{20}$Ne-emulsion, and 500 A MeV $^{56}$Fe-emulsion interactions. It can be seen that for interactions with $n_{h}\geq8$,  $<n_{g}^{f}>$ decreases with the increase of $n_{b}$ firstly and then becomes saturation except for 500 A MeV $^{56}$Fe-AgBr interaction, where $<n_{g}^{f}>$ decreases firstly and then increases with the increase of $n_{b}$, and finally becomes saturation; $<n_{g}^{b}>$ decreases slowly with increase of $n_{b}$ except for 500 A MeV $^{56}$Fe-AgBr interaction where $<n_{g}^{b}>$ increases slowly with increase of $n_{b}$. For interactions with $2\leq{n_{h}}\leq7$, $<n_{g}^{f}>$ decreases with the increase of $n_{b}$, $<n_{g}^{b}>$ almost keep constant with the increase of $n_{b}$. For all of the interactions ($n_{h}\geq0$), $<n_{g}^{b}>$ increases slowly with the increase of $n_{b}$, and $<n_{g}^{f}>$ increases with the increase of $n_{b}$ firstly and then becomes saturation except for 150 A MeV $^{4}$He-emulsion interactions where events statistics is lower and the correlation is not obvious. The correlations can be represented by a linear relation of the formula:
\begin{equation}
<n_{g}^{j}>=a.n_{b}+b
\end{equation}
where {\em j} means forward hemisphere ({\em f}) and backward hemisphere ({\em b}) respectively. The fitted lines is shown in the Fig.2 and the fitting parameters is presented in tables 3-5, some of the fitting parameters is from the first a few data sets.

\begin{sidewaystable}
\begin{center}
Table 2. The averaged forward and backward multiplicity and forward-backward multiplicity ratio of black, grey and heavily ionized track particles in nucleus-emulsion interactions.\\
\begin{tabular}{lccccccccc}\hline
Type of interaction        & $<n_{b}^{b}>$ & $<n_{b}^{f}>$ & $<n_{g}^{b}>$ & $<n_{g}^{f}>$ & $<n_{h}^{b}>$ & $<n_{h}^{f}>$ &$(F/B)_{b}$& $(F/B)_{g}$& $(F/B)_{h}$  \\\hline
150 A MeV $^{4}$He-H    & $0.16\pm0.06$ & $0.26\pm0.07$ & $0.05\pm0.04$ & $0.34\pm0.08$ & $0.21\pm0.07$ & $0.61\pm0.08$ & 1.63 & 6.80 & 2.90 \\
150 A MeV $^{4}$He-CNO  & $0.71\pm0.05$ & $1.50\pm0.07$ & $0.31\pm0.04$ & $1.42\pm0.08$ & $1.01\pm0.06$ & $2.92\pm0.08$ & 2.11 & 4.58 & 2.89 \\
150 A MeV $^{4}$He-Em   & $0.89\pm0.06$ & $1.74\pm0.09$ & $0.37\pm0.04$ & $1.62\pm0.09$ & $1.25\pm0.07$ & $3.36\pm0.13$ & 1.96 & 4.38 & 2.69 \\
150 A MeV $^{4}$He-AgBr & $2.42\pm0.25$ & $4.14\pm0.36$ & $0.92\pm0.18$ & $3.66\pm0.43$ & $3.34\pm0.25$ & $7.80\pm0.45$ & 1.71 & 3.98 & 2.34 \\
290 A MeV $^{12}$C-H    & $0.07\pm0.01$ & $0.30\pm0.03$ & $0.03\pm0.01$ & $0.27\pm0.02$ & $0.10\pm0.02$ & $0.56\pm0.03$ & 4.29 & 9.00 & 5.60 \\
290 A MeV $^{12}$C-CNO  & $0.68\pm0.02$ & $1.91\pm0.04$ & $0.26\pm0.02$ & $1.37\pm0.04$ & $0.94\pm0.03$ & $3.28\pm0.04$ & 2.81 & 5.27 & 3.49 \\
290 A MeV $^{12}$C-Em   & $0.94\pm0.03$ & $2.43\pm0.05$ & $0.40\pm0.02$ & $1.69\pm0.04$ & $1.34\pm0.04$ & $4.12\pm0.08$ & 2.59 & 4.23 & 3.07 \\
290 A MeV $^{12}$C-AgBr & $2.28\pm0.07$ & $5.38\pm0.13$ & $1.02\pm0.05$ & $3.58\pm0.11$ & $3.30\pm0.08$ & $8.96\pm0.15$ & 2.36 & 3.51 & 2.72 \\
400 A MeV $^{12}$C-H    & $0.08\pm0.03$ & $0.16\pm0.04$ & $0.00\pm0.00$ & $0.31\pm0.05$ & $0.08\pm0.03$ & $0.47\pm0.05$ & 2.00 &      & 5.88 \\
400 A MeV $^{12}$C-CNO  & $0.98\pm0.04$ & $1.95\pm0.06$ & $0.10\pm0.01$ & $1.36\pm0.06$ & $1.08\pm0.05$ & $3.31\pm0.07$ & 1.99 & 13.60& 3.06 \\
400 A MeV $^{12}$C-Em   & $1.53\pm0.06$ & $3.09\pm0.11$ & $0.26\pm0.02$ & $1.85\pm0.06$ & $1.78\pm0.07$ & $4.94\pm0.14$ & 2.02 & 7.12 & 2.78 \\
400 A MeV $^{12}$C-AgBr & $3.08\pm0.12$ & $6.31\pm0.21$ & $0.64\pm0.06$ & $3.32\pm0.13$ & $3.72\pm0.13$ & $9.63\pm0.24$ & 2.05 & 5.19 & 2.59 \\
400 A MeV $^{20}$Ne-H   & $0.20\pm0.03$ & $0.34\pm0.04$ & $0.06\pm0.02$ & $0.07\pm0.02$ & $0.26\pm0.04$ & $0.41\pm0.04$ & 1.70 & 1.17 & 1.58 \\
400 A MeV $^{20}$Ne-CNO & $0.92\pm0.04$ & $1.63\pm0.05$ & $0.38\pm0.03$ & $0.90\pm0.04$ & $1.30\pm0.04$ & $2.53\pm0.05$ & 1.77 & 2.37 & 1.95 \\
400 A MeV $^{20}$Ne-Em  & $1.26\pm0.04$ & $2.35\pm0.07$ & $0.51\pm0.03$ & $1.18\pm0.04$ & $1.77\pm0.05$ & $3.54\pm0.09$ & 1.87 & 2.31 & 2.00 \\
400 A MeV $^{20}$Ne-AgBr& $2.62\pm0.09$ & $5.12\pm0.16$ & $1.07\pm0.07$ & $2.44\pm0.11$ & $3.69\pm0.10$ & $7.56\pm0.17$ & 1.95 & 2.28 & 2.05 \\
500 A MeV $^{56}$Fe-H   & $0.12\pm0.02$ & $0.32\pm0.03$ & $0.02\pm0.01$ & $0.30\pm0.03$ & $0.14\pm0.02$ & $0.62\pm0.03$ & 2.67 & 15.00& 4.43 \\
500 A MeV $^{56}$Fe-CNO & $0.69\pm0.03$ & $1.38\pm0.05$ & $0.14\pm0.02$ & $2.33\pm0.07$ & $0.83\pm0.04$ & $3.72\pm0.07$ & 2.00 & 16.64& 4.48 \\
500 A MeV $^{56}$Fe-Em  & $1.24\pm0.04$ & $2.87\pm0.09$ & $0.38\pm0.02$ & $4.86\pm0.15$ & $1.61\pm0.05$ & $7.73\pm0.22$ & 2.31 & 12.79& 4.80 \\
500 A MeV $^{56}$Fe-AgBr& $2.21\pm0.07$ & $5.32\pm0.15$ & $0.75\pm0.04$ & $9.13\pm0.25$ & $2.96\pm0.09$ & $14.46\pm0.33$& 2.41 & 12.17& 4.89 \\\hline
\end{tabular}
\end{center}
\end{sidewaystable}

Fig. 3 shows the correlation between $<n_{b}^{b}>$, $<n_{b}^{f}>$ and $n_{g}$ for different type of 150 A MeV $^{4}$He-emulsion, 290 A MeV $^{12}$C-emulsion, 400 A MeV $^{12}$C-emulsion, 400 A MeV $^{20}$Ne-emulsion, and 500 A MeV $^{56}$Fe-emulsion interactions. It can be seen that for interactions with $n_{h}\geq8$, $<n_{b}^{f}>$ decreases with the increase of $n_{g}$ firstly and then becomes saturation except for 500 A MeV $^{56}$Fe-AgBr interaction, where $<n_{b}^{f}>$ decreases firstly and then increases with the increase of $n_{b}$, and finally becomes saturation; $<n_{b}^{b}>$ decreases slowly with increase of $n_{g}$ except for 500 A MeV $^{56}$Fe-AgBr interaction where the same tendency is happened as the ones in the correlation of $<n_{b}^{f}>$ and $n_{g}$. For interactions with $2\leq{n_{h}}\leq7$, $<n_{b}^{f}>$ decreases with the increase of $n_{b}$, and $<n_{b}^{b}>$ decreases slowly with the increase of $n_{g}$. For all of the interactions, $<n_{b}^{b}>$ increases slowly with the increase of $n_{g}$ except for 150 A MeV $^{4}$He-emulsion and 500 A MeV $^{56}$Fe-emulsion interactions, where $<n_{b}^{b}>$ increases firstly and then becomes saturation with the increase of $n_{g}$ for $^{56}$Fe-emulsion and for $^{4}$He-emulsion interactions the fitted slope is $-0.005\pm0.030$,  and $<n_{b}^{f}>$ increases with the increase of $n_{g}$ firstly and then becomes saturation except for 150 A MeV $^{4}$He-emulsion interactions where the fitted slope is $-0.030\pm0.048$. The correlations can be represented by a linear relation which is the same as equ.(1). The fitted lines is shown in the Fig.3 and the fitting parameters is presented in tables 3-5, some of the fitting parameters is from the first a few data sets.

\begin{figure}[htbp]
\begin{center}
\includegraphics[width=1.0\linewidth]{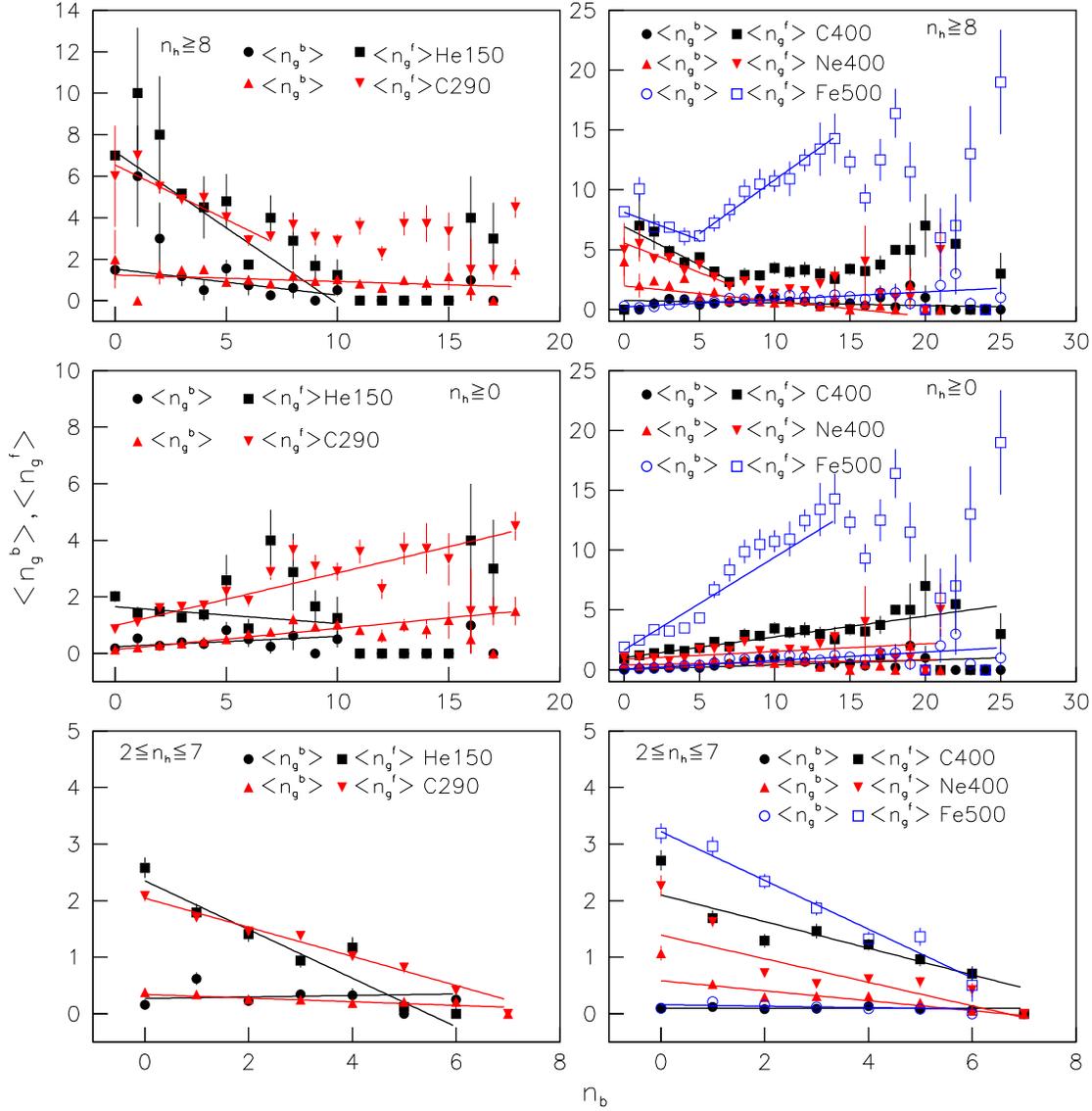}
\caption{(Color online) Correlations between the averaged multiplicity of backward and forward grey track particles and the number of black track particle for different type of interactions.}
\end{center}
\end{figure}

\begin{figure}[htbp]
\begin{center}
\includegraphics[width=1.0\linewidth]{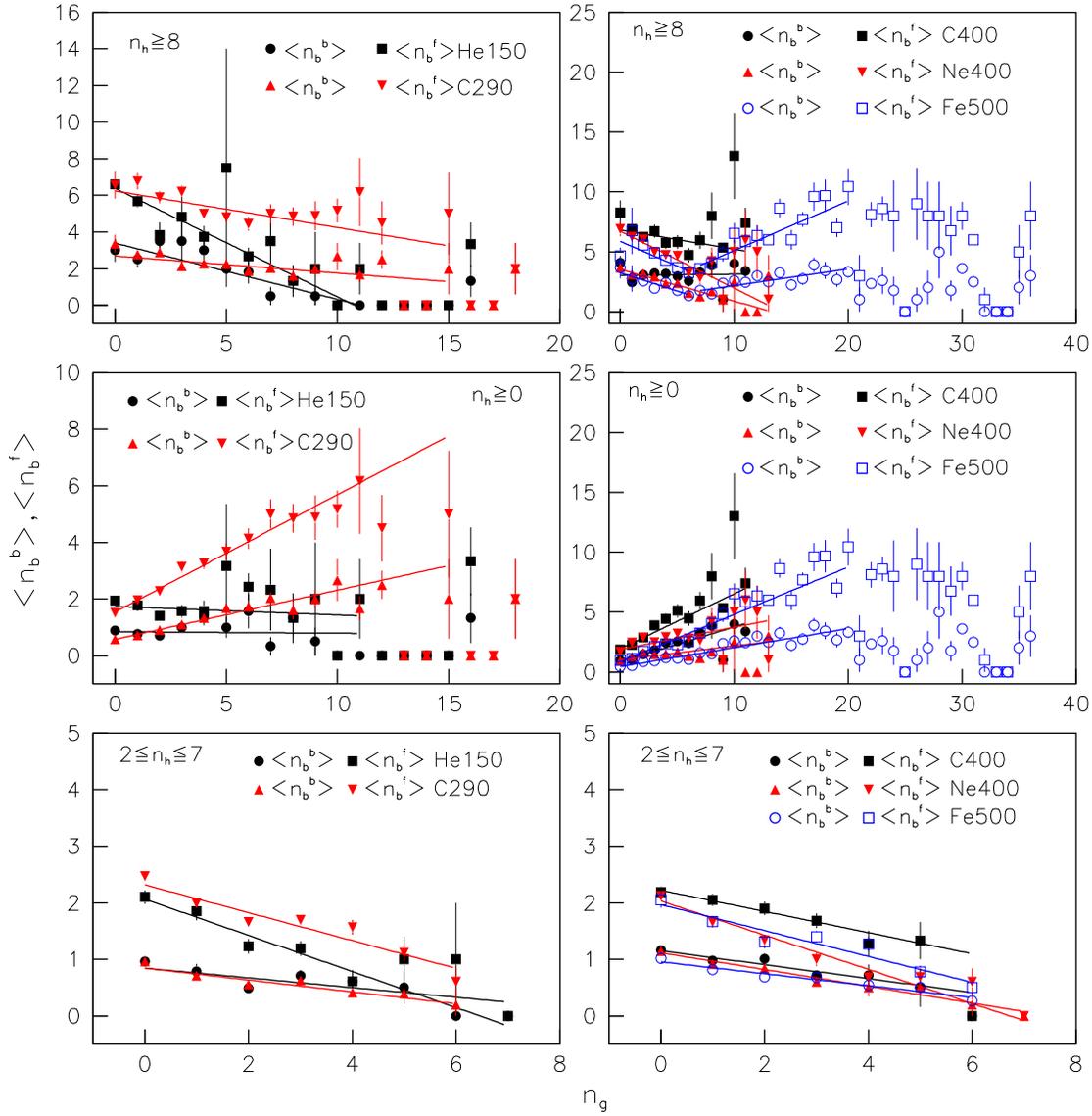}
\caption{(Color online) Correlations between the averaged multiplicity of backward and forward black track particles and the number of grey track particle for different type of interactions.}
\end{center}
\end{figure}

Fig. 4 shows the correlation between $<n_{b}^{b}>$, $<n_{b}^{f}>$, $<n_{g}^{b}>$, $<n_{g}^{f}>$ and $n_{h}$ for different type of 150 A MeV $^{4}$He-emulsion, 290 A MeV $^{12}$C-emulsion, 400 A MeV $^{12}$C-emulsion, 400 A MeV $^{20}$Ne-emulsion, and 500 A MeV $^{56}$Fe-emulsion interactions. It can be seen that for different type of interactions $<n_{b}^{f}>$ and $<n_{g}^{f}>$ increases with the increase of $n_{h}$, $<n_{b}^{b}>$ and $<n_{g}^{b}>$ increases slowly with the increase of $n_{h}$. The correlations can be represented by a linear relation which is the same as equ.(1). The fitted lines is shown in the Fig.4 and the fitting parameters is presented in tables 3-5.

\begin{figure}[htbp]
\begin{center}
\includegraphics[width=1.0\linewidth]{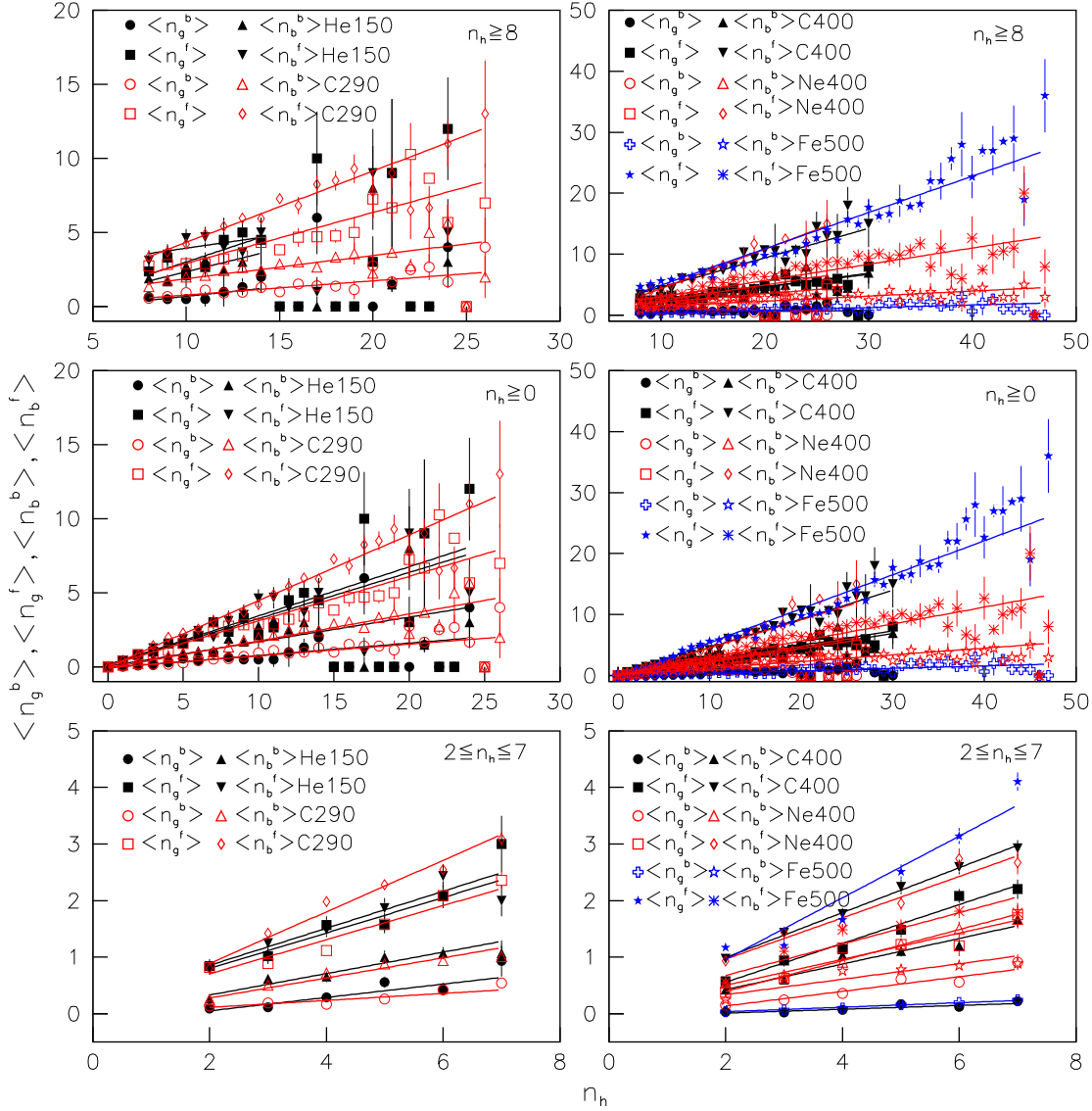}
\caption{(Color online) Correlations between the averaged multiplicity of backward-forward black and grey track particles and the number of heavily ionized track particle for different type of interactions.}
\end{center}
\end{figure}

Fig. 5 shows the correlation between $<n_{b}^{f}>$, $<n_{g}^{b}>$, $<n_{g}^{f}>$, $<n_{h}^{b}>$, $<n_{h}^{f}>$ and $n_{b}^{b}$ for different type of 150 A MeV $^{4}$He-emulsion, 290 A MeV $^{12}$C-emulsion, 400 A MeV $^{12}$C-emulsion, 400 A MeV $^{20}$Ne-emulsion, and 500 A MeV $^{56}$Fe-emulsion interactions. For interactions with $2\leq{n_{h}}\leq7$, $<n_{h}^{b}>$ and $<n_{g}^{f}>$ increases with the increase of $n_{b}^{b}$, $<n_{h}^{f}>$ and $<n_{g}^{b}>$ decreases with the increase of $n_{b}^{b}$ except for the correlation of $<n_{g}^{b}>$ and $n_{b}^{b}$ for 400 A MeV $^{12}$C where the fitted slope parameter is $0.001\pm0.018$, $<n_{b}^{f}>$ changes slowly with the increase of $n_{b}^{b}$ and the error of fitted slope parameter is big. For interactions with $n_{h}\geq8$, $<n_{h}^{f}>$, $<n_{h}^{b}>$, $<n_{g}^{f}>$, and $<n_{b}^{f}>$ increase with the increase of $n_{b}^{b}$ except for case of 150 A MeV $^{4}$He where the events statistics is lower and the error of fitted slope parameter is big, and for the correlation of $<n_{h}^{f}>$ and $n_{b}^{b}$ for case of 400 A MeV $^{20}$Ne where the fitted slope parameter is $-0.028\pm0.117$, $<n_{g}^{b}>$ decreases with the increase of $n_{b}^{b}$ except for the case of 500 A MeV $^{56}$Fe where $<n_{g}^{b}>$ increases with the increase of $n_{b}^{b}$. For all of the interactions, $<n_{b}^{f}>$, $<n_{g}^{b}>$, $<n_{g}^{f}>$, $<n_{h}^{b}>$, and $<n_{h}^{f}>$ increase with the increase of  $n_{b}^{b}$ except for correlation of $<n_{g}^{b}>$ and $n_{b}^{b}$ for case of 150 A MeV $^{4}$He where the fitted slope is $-0.015\pm0.037$. The correlations can be represented by a linear relation which is the same as equ.(1). The fitted lines is shown in the Fig.5 and the fitting parameters is presented in tables 3-5.

\begin{figure}[htbp]
\begin{center}
\includegraphics[width=1.0\linewidth]{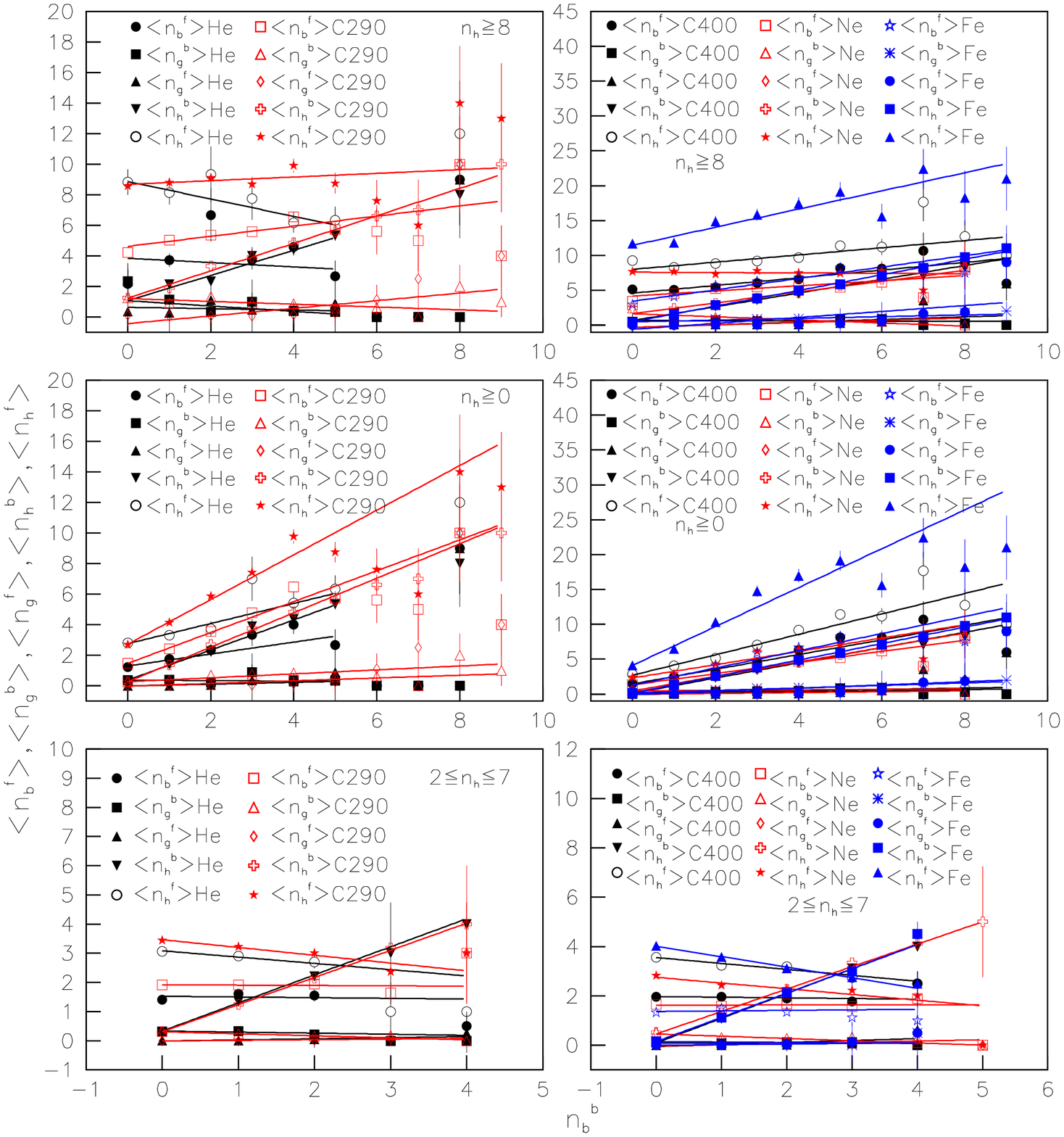}
\caption{(Color online) Correlations between $<n_{b}^{f}>$, $<n_{g}^{b}>$, $<n_{g}^{f}>$, $<n_{h}^{b}>$, and $<n_{h}^{f}>$ and $n_{b}^{b}$ for different type of interactions.}
\end{center}
\end{figure}

\begin{figure}[htbp]
\begin{center}
\includegraphics[width=1.0\linewidth]{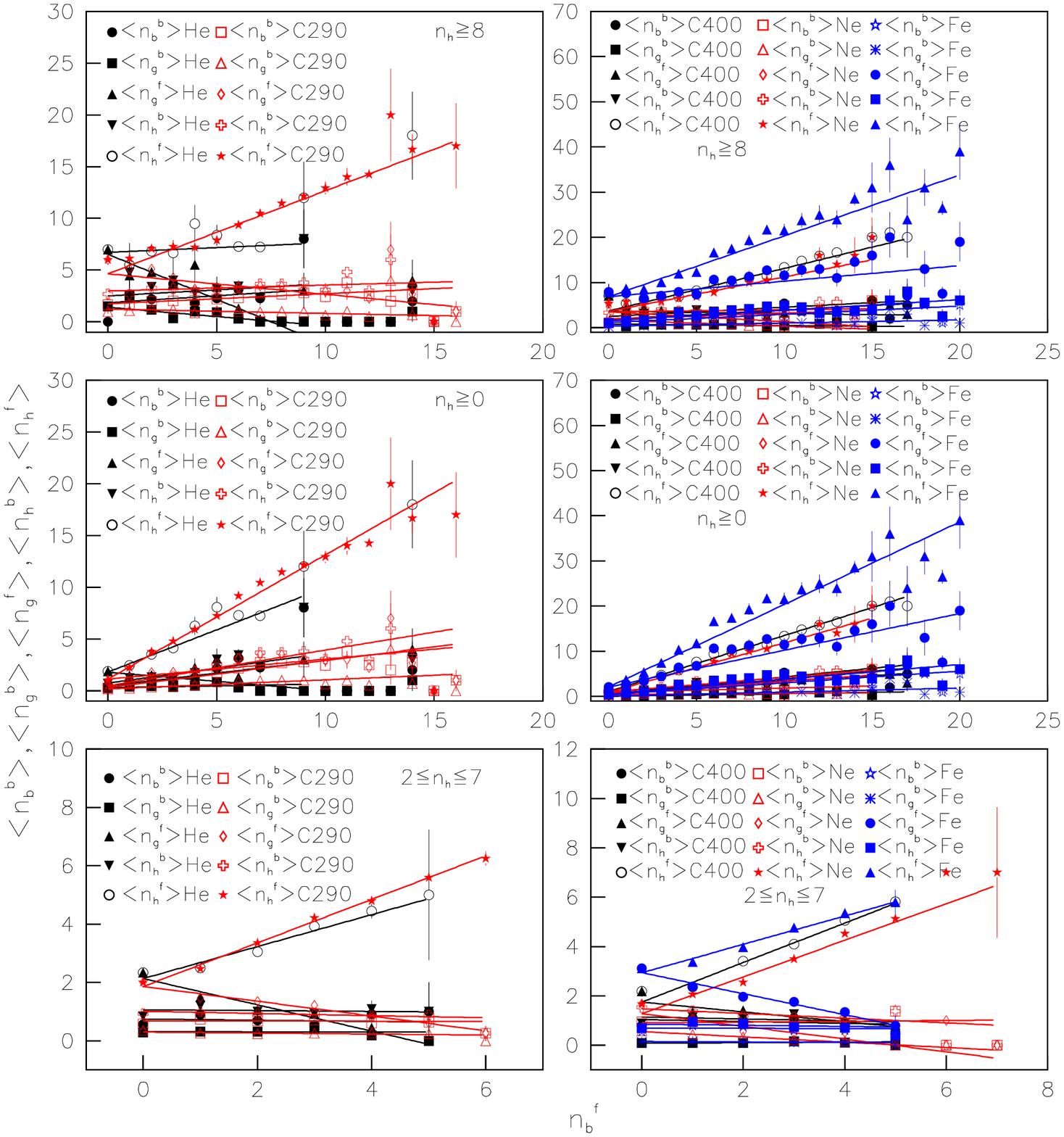}
\caption{(Color online) Correlations between $<n_{b}^{b}>$, $<n_{g}^{b}>$, $<n_{g}^{f}>$, $<n_{h}^{b}>$, and $<n_{h}^{f}>$ and $n_{b}^{f}$ for different type of interactions.}
\end{center}
\end{figure}

Fig. 6 shows the correlation between $<n_{b}^{b}>$, $<n_{g}^{b}>$, $<n_{g}^{f}>$, $<n_{h}^{b}>$, $<n_{h}^{f}>$ and $n_{b}^{f}$ for different type of 150 A MeV $^{4}$He-emulsion, 290 A MeV $^{12}$C-emulsion, 400 A MeV $^{12}$C-emulsion, 400 A MeV $^{20}$Ne-emulsion, and 500 A MeV $^{56}$Fe-emulsion interactions. For interactions with $2\leq{n_{h}}\leq7$, $<n_{h}^{b}>$,  $<n_{g}^{b}>$, $<n_{g}^{f}>$ and $<n_{b}^{b}>$ decrease with the increase of $n_{b}^{f}$ except for correlation of $<n_{g}^{b}>$ and $n_{b}^{f}$ for case of 400 A MeV $^{12}$C where the fitted slope is $0.011\pm0.013$ and for correlation of $<n_{b}^{b}>$ and $n_{b}^{f}$ for case of 400 A MeV $^{20}$Ne where fitted slope is $0.016\pm0.030$, $<n_{h}^{f}>$ increases with the increase of $n_{b}^{f}$. For interactions with $n_{h}\geq8$, $<n_{h}^{f}>$, $<n_{h}^{b}>$, and $<n_{b}^{b}>$ increase with the increase of $n_{b}^{f}$, $<n_{g}^{f}>$ and $<n_{g}^{b}>$ decreases with the increase of $n_{b}^{f}$ except for the case of 500 A MeV $^{56}$Fe where $<n_{g}^{f}>$ and $<n_{g}^{b}>$ increases with the increase of $n_{b}^{f}$. For all of the interactions, $<n_{b}^{b}>$, $<n_{g}^{b}>$, $<n_{g}^{f}>$, $<n_{h}^{b}>$, and $<n_{h}^{f}>$ increase with the increase of  $n_{b}^{f}$ except for correlation of $<n_{g}^{f}>$ and $n_{b}^{f}$ for case of 150 A MeV $^{4}$He where the fitted slope is $-0.117\pm0.036$ and changing tendency is not obvious. The correlations can be represented by a linear relation which is the same as equ.(1). The fitted lines is shown in the Fig.6 and the fitting parameters is presented in tables 3-5.

\begin{figure}[htbp]
\begin{center}
\includegraphics[width=1.0\linewidth]{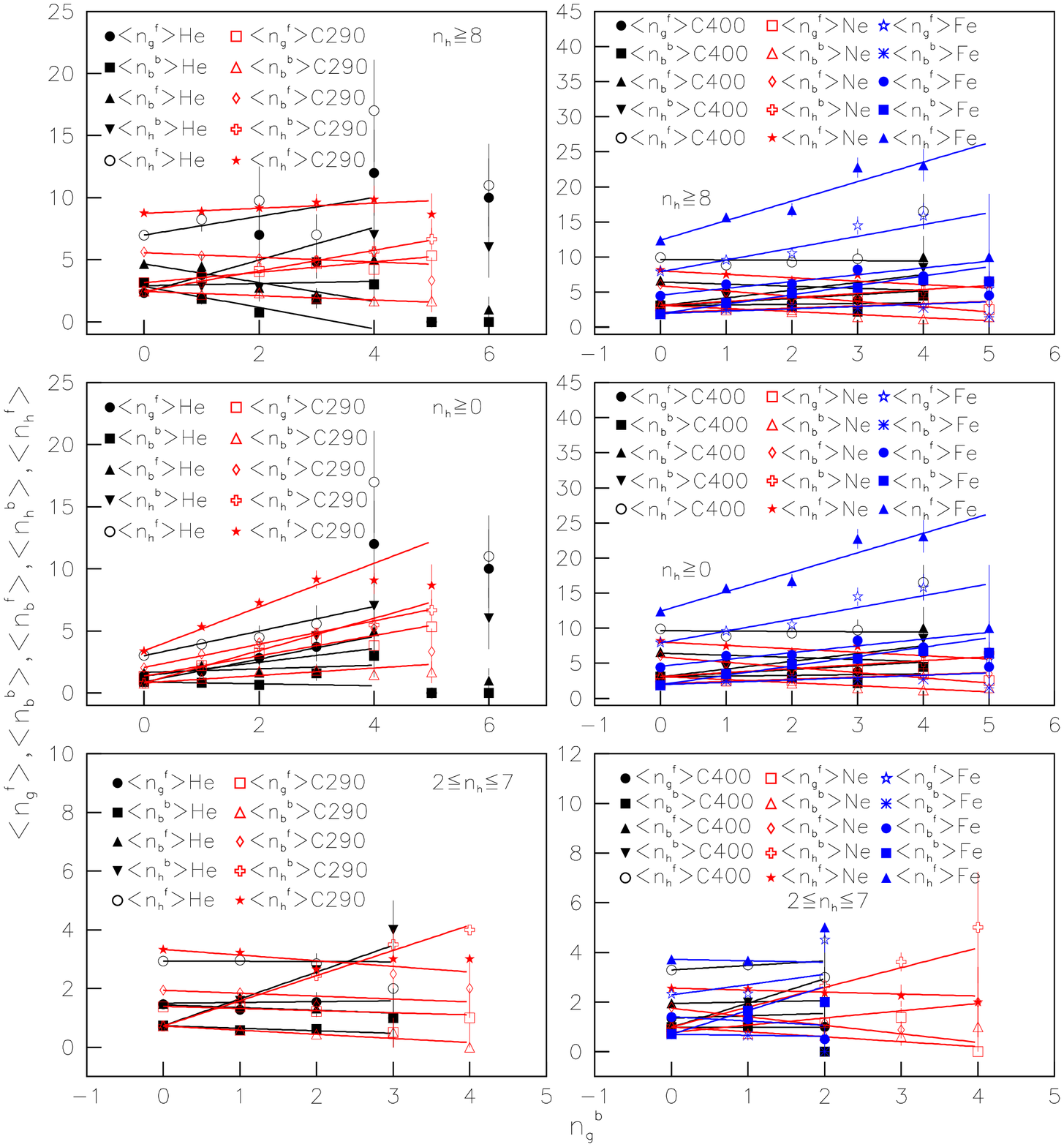}
\caption{(Color online) Correlations between $<n_{g}^{f}>$, $<n_{b}^{b}>$, $<n_{b}^{f}>$, $<n_{h}^{b}>$, and $<n_{h}^{f}>$ and $n_{g}^{b}$ for different type of interactions.}
\end{center}
\end{figure}

\begin{figure}[htbp]
\begin{center}
\includegraphics[width=1.0\linewidth]{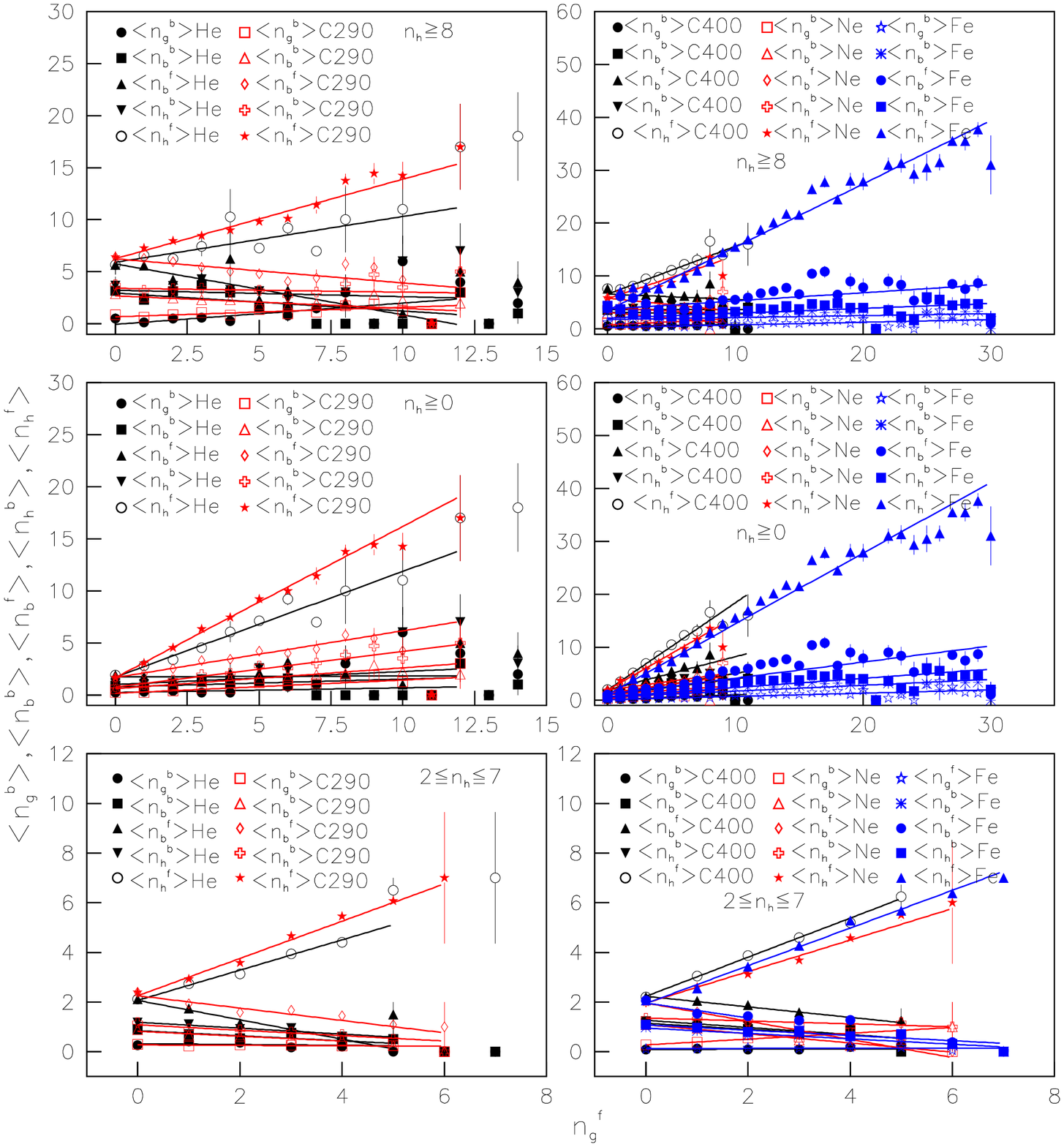}
\caption{(Color online) Correlations between $<n_{g}^{b}>$, $<n_{b}^{b}>$, $<n_{b}^{f}>$, $<n_{h}^{b}>$, and $<n_{h}^{f}>$ and $n_{g}^{f}$ for different type of interactions.}
\end{center}
\end{figure}

Fig. 7 shows the correlation between $<n_{g}^{f}>$, $<n_{b}^{b}>$, $<n_{b}^{f}>$, $<n_{h}^{b}>$, $<n_{h}^{f}>$ and $n_{g}^{b}$ for different type of 150 A MeV $^{4}$He-emulsion, 290 A MeV $^{12}$C-emulsion, 400 A MeV $^{12}$C-emulsion, 400 A MeV $^{20}$Ne-emulsion, and 500 A MeV $^{56}$Fe-emulsion interactions. For interactions with $2\leq{n_{h}}\leq7$, $<n_{h}^{f}>$,  $<n_{b}^{b}>$, and $<n_{b}^{f}>$ decrease with the increase of $n_{g}^{b}$ except for correlation of $<n_{h}^{f}>$ and $n_{g}^{b}$ for case of 400 A MeV $^{12}$C where the fitted slope is $0.180\pm0.215$ and for correlation of $<n_{b}^{f}>$ and $n_{g}^{b}$ for cases of 150 A MeV $^{4}$He and 400 A MeV $^{20}$Ne where fitted slope are $0.027\pm0.114$ and $0.058\pm0.184$ respectively, $<n_{h}^{b}>$ and $<n_{g}^{f}>$ increase with the increase of $n_{g}^{b}$ except for the correlation of $<n_{g}^{f}>$ and $n_{g}^{b}$ for cases of 150 A MeV $^{4}$He and 290 A MeV $^{12}$C where the error of fitted slope is big. For interactions with $n_{h}\geq8$, $<n_{h}^{f}>$, $<n_{h}^{b}>$, and $<n_{g}^{f}>$ increase with the increase of $n_{g}^{b}$ except for correlation of $<n_{h}^{f}>$ and $n_{g}^{b}$ for cases of 400 A MeV $^{20}$Ne and 400 A MeV $^{12}$C where the fitted slope is negative, $<n_{b}^{f}>$ and $<n_{b}^{b}>$ decreases with the increase of $n_{g}^{b}$ except for the case of 500 A MeV $^{56}$Fe where $<n_{b}^{f}>$ and $<n_{b}^{b}>$ increases with the increase of $n_{g}^{b}$ and for the correlation of $<n_{b}^{b}>$ and $n_{g}^{b}$ where the fitted slope is positive. For all of the interactions, $<n_{g}^{f}>$, $<n_{b}^{b}>$, $<n_{b}^{f}>$, $<n_{h}^{b}>$, and $<n_{h}^{f}>$ increase with the increase of $n_{g}^{b}$ except for correlation of $<n_{b}^{b}>$ and $n_{g}^{b}$ for case of 150 A MeV $^{4}$He where the fitted slope is $-0.078\pm0.064$ and changing tendency is not obvious. The correlations can be represented by a linear relation which is the same as equ.(1). The fitted lines is shown in the Fig.7 and the fitting parameters is presented in tables 3-5.

Fig. 8 shows the correlation between $<n_{g}^{b}>$, $<n_{b}^{b}>$, $<n_{b}^{f}>$, $<n_{h}^{b}>$, $<n_{h}^{f}>$ and $n_{g}^{f}$ for different type of 150 A MeV $^{4}$He-emulsion, 290 A MeV $^{12}$C-emulsion, 400 A MeV $^{12}$C-emulsion, 400 A MeV $^{20}$Ne-emulsion, and 500 A MeV $^{56}$Fe-emulsion interactions. For interactions with $2\leq{n_{h}}\leq7$, $<n_{h}^{b}>$,  $<n_{b}^{b}>$, and $<n_{b}^{f}>$ decrease with the increase of $n_{g}^{f}$, $<n_{h}^{f}>$ and $<n_{g}^{b}>$ increase with the increase of $n_{g}^{f}$ except for the correlation of $<n_{g}^{b}>$ and $n_{g}^{f}$ for cases of 150 A MeV $^{4}$He and 290 A MeV $^{12}$C where the fitted slope are $-0.016\pm0.029$ and $-0.008\pm0.014$ respectively. For interactions with $n_{h}\geq8$, $<n_{h}^{f}>$ and $<n_{g}^{b}>$ increase with the increase of $n_{g}^{f}$, $<n_{h}^{b}>$, $<n_{b}^{b}>$, and $<n_{b}^{f}>$ decreases with the increase of $n_{g}^{f}$ except for the case of 500 A MeV $^{56}$Fe where the fitted slope is positive and for the correlation of $<n_{h}^{b}>$ and $n_{g}^{f}$ for case of 400 A MeV $^{12}$C where the fitted slope is $0.001\pm0.066$. For all of the interactions, $<n_{g}^{b}>$, $<n_{b}^{b}>$, $<n_{b}^{f}>$, $<n_{h}^{b}>$, and $<n_{h}^{f}>$ increase with the increase of $n_{g}^{f}$. The correlations can be represented by a linear relation which is the same as equ.(1). The fitted lines is shown in the Fig.8 and the fitting parameters is presented in tables 3-5.

Based on geometrical picture (the participant-spectator model)\cite{bowm} and the cascade evaporation model\cite{powe} of high nucleus-nucleus collisions, the grey track particles are emitted from the target nucleus very soon after the instant of impact, which is the target recoiled protons of energy ranging up to 400 MeV, the black track particles are images of target evaporated particles of low-energy($E<30$MeV) and singly or multiply charged fragments. For interactions with $2\leq{n_{h}}\leq7$ (nucleus-CHO interactions), $<n_{g}^{b}>$ and $<n_{g}^{f}>$ decrease with increase of $n_{b}$, $<n_{b}^{b}>$ and $<n_{b}^{f}>$ also decrease with increase of $n_{g}$ because of the limited target size (the maximum target fragments $n_{h}=8$ which corresponds to total disintegration of oxygen nucleus). For interactions with $n_{h}\geq8$ (nucleus-AgBr interactions) and the light projectile (less than $^{20}$Ne) $<n_{g}^{b}>$ decreases with increase of $n_{b}$, and $<n_{g}^{f}>$ decreases firstly and then keep constant with increase of $n_{b}$ because the minimum target fragments is $n_{h}=8$ and excitation energy of target residues increases with increase of the number of cascading collisions but it is limited for limited projectile. For interactions with $n_{h}\geq8$ and heavy projectile ($^{56}$Fe projectile), with increasing of $n_{b}$ the excitation energy of target residues and the number of cascading collisions is increased, $<n_{g}^{f}>$ decreases firstly and then increases with increase of $n_{b}$ and finally becomes saturation because of limited target size. The same mechanism can be used to explain the correlations of $<n_{b}^{b}>$, $<n_{b}^{f}>$ and $n_{g}$ for interactions with $n_{h}\geq8$. For all of the interactions with increasing of $n_{b}$ the averaged excitation energy of target residues and the cascading collision in target spectator increased, $<n_{g}^{b}>$ and $<n_{g}^{f}>$ will increase but due to limited target size $<n_{g}^{f}>$ finally becomes saturation, $<n_{b}^{f}>$ and $<n_{b}^{b}>$ increase with increase of $n_{g}$ and for interactions of 500 A MeV $^{56}$Fe-emulsion which become saturation when $n_{g}$ is greater than 20. With increasing of $n_{h}$ the excitation energy of target residues and the number of cascading collisions increased, so $<n_{b}^{b}>$, $<n_{b}^{f}>$, $<n_{g}^{b}>$, and $<n_{g}^{f}>$ will increase with increase of $n_{h}$.

The backward-forward multiplicity correlations can also be explained based on geometrical picture\cite{bowm} and the cascade evaporation model\cite{powe} of high nucleus-nucleus collisions. For the interactions with $2\leq{n_{h}}\leq7$, with increasing of $n_{b}^{b}$ the excitation energy of target residues and the cascading collision in target spectator increased, $<n_{h}^{b}>$ and $<n_{g}^{f}>$ increased,  but $<n_{h}^{f}>$,  $<n_{g}^{b}>$, and $<n_{b}^{f}>$ decreased because of the limited target spectators. For the same reasons with increasing of $n_{b}^{f}$, $<n_{h}^{f}>$ increased but $<n_{h}^{b}>$,  $<n_{g}^{b}>$, $<n_{g}^{b}>$, and $<n_{b}^{b}>$ decreased. With increasing of $n_{g}^{b}$, $<n_{h}^{b}>$ and $<n_{g}^{f}>$ increased, but $<n_{h}^{f}>$, $<n_{g}^{b}>$, and $<n_{b}^{f}>$ decreased. With increasing of $n_{g}^{f}$, $<n_{g}^{b}>$ and $<n_{h}^{f}>$ increased but $<n_{h}^{b}>$,  $<n_{b}^{b}>$, and $<n_{b}^{f}>$ decreased. For interactions with $n_{h}\geq8$ and light projectile (less than $^{20}$Ne), with increasing of $n_{b}^{b}$, $<n_{g}^{b}>$ decreased but $<n_{h}^{f}>$,  $<n_{h}^{b}>$, $<n_{g}^{f}>$ and $<n_{b}^{f}>$ increased; with increasing of $n_{b}^{f}$, $<n_{g}^{f}>$ and $<n_{g}^{b}>$ decreased but $<n_{h}^{b}>$,  $<n_{h}^{f}>$, and $<n_{b}^{b}>$ increased; with increasing of $n_{g}^{b}$, $<n_{b}^{f}>$ and $<n_{b}^{b}>$ decreased but $<n_{h}^{f}>$,  $<n_{h}^{b}>$, and $<n_{g}^{f}>$ increased; with increasing of $n_{g}^{f}$, $<n_{b}^{f}>$, $<n_{b}^{b}>$, and $<n_{h}^{b}>$ decreased but $<n_{h}^{f}>$ and $<n_{g}^{b}>$ increased. For interactions with $n_{h}\geq8$ and heavy projectile ($^{56}$Fe), the slopes of all of the backward-forward multiplicity correlations are positive because of the number of cascading collisions is greater and the excitation energy of target residues is greater. For all of the interactions the slopes of all of the backward-forward multiplicity correlations are positive.

\begin{sidewaystable}
Table 3. The fitting parameters of forward-backward multiplicity correlations for nucleus-emulsion interactions with $2\leq{n_{h}}\leq7$ using linear relation Eq.(1).
\begin{small}
\begin{tabular}{lcccccccccc}\hline
Correlation     & \multicolumn{2}{c}{150 A MeV $^{4}$He} & \multicolumn{2}{c}{290 A MeV $^{12}$C} & \multicolumn{2}{c}{400 A MeV $^{12}$C} & \multicolumn{2}{c}{400 A MeV $^{20}$Ne} & \multicolumn{2}{c}{500 A MeV $^{56}$Fe} \\
                          &      a           &    b            &       a          &      b          &       a          &        b        &       a       &       b        &       a       &      b    \\\hline
$<n_{g}^{b}>-n_{b}$       & $0.01\pm0.03$  & $0.27\pm0.07$ & $-0.03\pm0.01$ & $0.34\pm0.04$ & $0.001\pm0.010$ & $0.10\pm0.03$ & $-0.09\pm0.01$ & $0.58\pm0.05$  & $-0.01\pm0.01$ & $0.16\pm0.03$    \\
$<n_{g}^{f}>-n_{b}$       & $-0.43\pm0.04$ & $2.35\pm0.11$ & $-0.26\pm0.02$ & $2.04\pm0.06$ & $-0.24\pm0.03$ & $2.10\pm0.10$ & $-0.21\pm0.02$ & $1.39\pm0.08$  & $-0.43\pm0.04$ & $3.22\pm0.18$    \\
$<n_{b}^{b}>-n_{g}$       & $-0.09\pm0.03$ & $0.84\pm0.07$ & $-0.11\pm0.02$ & $0.85\pm0.04$ & $-0.12\pm0.03$ & $1.15\pm0.07$ & $-0.15\pm0.02$ & $1.12\pm0.05$  & $-0.11\pm0.02$ & $0.96\pm0.06$        \\
$<n_{b}^{f}>-n_{g}$       & $-0.32\pm0.04$ & $2.06\pm0.10$ & $-0.25\pm0.02$ & $2.32\pm0.06$ & $-0.19\pm0.04$ & $2.22\pm0.08$ & $-0.30\pm0.02$ & $2.30\pm0.06$  & $-0.23\pm0.02$ & $1.97\pm0.08$          \\
$<n_{g}^{b}>-n_{h}$       & $0.12\pm0.02$  & $-0.19\pm0.07$& $0.06\pm0.01$  & $-0.01\pm0.04$& $0.03\pm0.01$  & $-0.06\pm0.03$& $0.13\pm0.02$  & $-0.12\pm0.05$ & $0.04\pm0.01$  & $-0.04\pm0.03$     \\
$<n_{g}^{f}>-n_{h}$       & $0.31\pm0.05$  & $0.17\pm0.18$ & $0.30\pm0.02$  & $0.10\pm0.07$ & $0.34\pm0.03$  & $-0.13\pm0.11$& $0.27\pm0.03$  & $-0.16\pm0.09$ & $0.54\pm0.03$  & $-0.13\pm0.12$      \\
$<n_{b}^{b}>-n_{h}$       & $0.19\pm0.03$  & $-0.05\pm0.10$& $0.18\pm0.01$  & $-0.08\pm0.05$& $0.23\pm0.03$  & $-0.03\pm0.09$& $0.23\pm0.02$  & $0.04\pm0.08$  & $0.14\pm0.02$  & $0.06\pm0.07$    \\
$<n_{b}^{f}>-n_{h}$       & $0.33\pm0.04$  & $0.20\pm0.16$ & $0.46\pm0.02$  & $-0.02\pm0.07$& $0.40\pm0.02$  & $0.20\pm0.11$ & $0.37\pm0.03$  & $0.23\pm0.10$  & $0.28\pm0.02$  & $0.12\pm0.10$  \\
$<n_{b}^{f}>-n_{b}^{b}$   & $-0.02\pm0.09$ & $1.52\pm0.10$ & $-0.01\pm0.05$ & $1.92\pm0.05$ & $-0.03\pm0.06$ & $1.97\pm0.09$ & $0.01\pm0.05$  & $1.62\pm0.07$  & $0.02\pm0.06$  & $1.37\pm0.07$  \\
$<n_{g}^{b}>-n_{b}^{b}$   & $-0.04\pm0.05$ & $0.33\pm0.05$ & $-0.07\pm0.02$ & $0.31\pm0.03$ & $0.001\pm0.018$  & $0.09\pm0.02$ & $-0.09\pm0.02$ & $0.46\pm0.04$  & $-0.02\pm0.02$ & $0.15\pm0.02$       \\
$<n_{g}^{f}>-n_{b}^{b}$   & $0.04\pm0.13$  & $-0.01\pm0.16$& $0.02\pm0.06$  & $-0.01\pm0.07$& $0.08\pm0.07$  & $-0.05\pm0.13$& $0.05\pm0.03$  & $-0.04\pm0.07$ & $0.03\pm0.14$  & $-0.01\pm0.18$         \\
$<n_{h}^{b}>-n_{b}^{b}$   & $0.96\pm0.05$  & $0.33\pm0.05$ & $0.93\pm0.02$  & $0.31\pm0.03$ & $1.00\pm0.02$  & $0.09\pm0.02$ & $0.91\pm0.02$  & $0.46\pm0.04$  & $0.98\pm0.02$  & $0.15\pm0.02$  \\
$<n_{h}^{f}>-n_{b}^{b}$   & $-0.22\pm0.11$ & $3.08\pm0.10$ & $-0.27\pm0.05$ & $3.47\pm0.06$ & $-0.24\pm0.07$ & $3.55\pm0.10$ & $-0.23\pm0.05$ & $2.75\pm0.07$  & $-0.43\pm0.07$ & $4.11\pm0.10$  \\
$<n_{b}^{b}>-n_{b}^{f}$   & $-0.02\pm0.04$ & $0.74\pm0.08$ & $-0.002\pm0.018$ & $0.68\pm0.04$ & $-0.04\pm0.03$ & $1.04\pm0.08$ & $0.02\pm0.03$  & $0.90\pm0.06$  & $-0.01\pm0.03$ & $0.70\pm0.05$  \\
$<n_{g}^{b}>-n_{b}^{f}$   & $-0.003\pm0.030$ & $0.31\pm0.06$ & $-0.02\pm0.01$ & $0.29\pm0.03$ & $0.01\pm0.01$  & $0.08\pm0.03$ & $-0.11\pm0.02$ & $0.55\pm0.05$  & $-0.01\pm0.01$ & $0.15\pm0.03$  \\
$<n_{g}^{f}>-n_{b}^{f}$   & $-0.45\pm0.05$ & $2.13\pm0.11$ & $-0.25\pm0.02$ & $1.86\pm0.06$ & $-0.20\pm0.04$ & $1.75\pm0.10$ & $-0.26\pm0.03$ & $1.28\pm0.07$  & $-0.43\pm0.05$ & $2.94\pm0.11$ \\
$<n_{h}^{b}>-n_{b}^{f}$   & $-0.02\pm0.05$ & $1.06\pm0.10$ & $-0.04\pm0.02$ & $1.01\pm0.05$ & $-0.04\pm0.04$ & $1.14\pm0.09$ & $-0.10\pm0.03$ & $1.48\pm0.07$  & $-0.02\pm0.03$ & $0.86\pm0.06$ \\
$<n_{h}^{f}>-n_{b}^{f}$   & $0.55\pm0.05$  & $2.13\pm0.11$ & $0.75\pm0.02$  & $1.86\pm0.06$ & $0.80\pm0.04$  & $1.75\pm0.10$ & $0.74\pm0.03$  & $1.28\pm0.07$  & $0.57\pm0.05$  & $2.94\pm0.11$ \\
$<n_{g}^{f}>-n_{g}^{b}$   & $-0.09\pm0.12$ & $1.44\pm0.09$ & $-0.07\pm0.06$ & $1.39\pm0.04$ & $0.10\pm0.19$  & $1.35\pm0.06$ & $0.29\pm0.06$  & $0.78\pm0.04$  & $0.42\pm0.16$  & $2.29\pm0.08$ \\
$<n_{b}^{b}>-n_{g}^{b}$   & $-0.09\pm0.06$ & $0.73\pm0.05$ & $-0.14\pm0.04$ & $0.72\pm0.03$ & $-0.001\pm0.158$ & $0.98\pm0.05$ & $-0.20\pm0.05$ & $1.00\pm0.04$  & $-0.04\pm0.11$ & $0.70\pm0.04$ \\
$<n_{b}^{f}>-n_{g}^{b}$   & $0.03\pm0.11$  & $1.50\pm0.08$ & $-0.10\pm0.07$ & $1.94\pm0.04$ & $0.06\pm0.18$  & $1.94\pm0.06$ & $-0.34\pm0.06$ & $1.75\pm0.06$  & $-0.17\pm0.12$ & $1.41\pm0.05$ \\
$<n_{h}^{b}>-n_{g}^{b}$   & $0.91\pm0.06$  & $0.73\pm0.05$ & $0.86\pm0.04$  & $0.72\pm0.03$ & $0.98\pm0.16$  & $0.98\pm0.05$ & $0.80\pm0.05$  & $1.00\pm0.04$  & $0.96\pm0.11$  & $0.70\pm0.04$  \\
$<n_{h}^{f}>-n_{g}^{b}$   & $-0.01\pm0.14$ & $2.94\pm0.09$ & $-0.19\pm0.09$ & $3.33\pm0.05$ & $0.18\pm0.22$  & $3.29\pm0.07$ & $-0.08\pm0.07$ & $2.56\pm0.06$  & $-0.06\pm0.21$ & $3.72\pm0.08$ \\
$<n_{g}^{b}>-n_{g}^{f}$   & $-0.02\pm0.03$ & $0.32\pm0.05$ & $-0.01\pm0.01$ & $0.27\pm0.03$ & $0.001\pm0.011$  & $0.09\pm0.02$ & $0.12\pm0.03$  & $0.27\pm0.03$  & $0.003\pm0.009$  & $0.12\pm0.03$ \\
$<n_{b}^{b}>-n_{g}^{f}$   & $-0.10\pm0.03$ & $0.84\pm0.07$ & $-0.10\pm0.02$ & $0.82\pm0.07$ & $-0.12\pm0.04$ & $1.14\pm0.07$ & $-0.18\pm0.03$ & $1.09\pm0.05$  & $-0.11\pm0.02$ & $0.95\pm0.06$ \\
$<n_{b}^{f}>-n_{g}^{f}$   & $-0.39\pm0.04$ & $2.07\pm0.09$ & $-0.25\pm0.03$ & $2.25\pm0.05$ & $-0.21\pm0.04$ & $2.23\pm0.08$ & $-0.37\pm0.03$ & $1.96\pm0.06$  & $-0.24\pm0.02$ & $1.94\pm0.08$ \\
$<n_{h}^{b}>-n_{g}^{f}$   & $-0.12\pm0.04$ & $1.19\pm0.09$ & $-0.11\pm0.02$ & $1.08\pm0.04$ & $-0.14\pm0.03$ & $1.26\pm0.07$ & $-0.06\pm0.03$ & $1.35\pm0.05$  & $-0.11\pm0.02$ & $1.08\pm0.07$ \\
$<n_{h}^{f}>-n_{g}^{f}$   & $0.61\pm0.04$  & $2.06\pm0.09$ & $0.75\pm0.03$  & $2.25\pm0.05$ & $0.79\pm0.04$  & $2.23\pm0.08$ & $0.63\pm0.03$  & $1.96\pm0.06$  & $0.76\pm0.02$  & $1.94\pm0.08$ \\\hline
\end{tabular}
\end{small}
\end{sidewaystable}

\begin{sidewaystable}
Table 4. The fitting parameters of forward-backward multiplicity correlations for nucleus-emulsion interactions with $n_{h}\geq8$ using linear relation Eq.(1).
\begin{small}
\begin{tabular}{lcccccccccc}\hline
Correlation     & \multicolumn{2}{c}{150 A MeV $^{4}$He} & \multicolumn{2}{c}{290 A MeV $^{12}$C} & \multicolumn{2}{c}{400 A MeV $^{12}$C} & \multicolumn{2}{c}{400 A MeV $^{20}$Ne} & \multicolumn{2}{c}{500 A MeV $^{56}$Fe} \\
                          &      a           &    b            &       a          &      b          &       a          &        b        &       a       &       b        &       a       &       b      \\\hline
$<n_{g}^{b}>-n_{b}$       & $-0.13\pm0.05$ & $1.53\pm0.33$ & $-0.03\pm0.02$ & $1.24\pm0.14$ & $-0.02\pm0.01$ & $0.76\pm0.13$ & $-0.13\pm0.02$ & $1.97\pm0.16$  & $0.06\pm0.01$  & $0.22\pm0.05$    \\
$<n_{g}^{f}>-n_{b}$       & $-0.73\pm0.06$ & $7.18\pm0.32$ & $-0.53\pm0.07$ & $6.54\pm0.35$ & $-0.64\pm0.11$ & $6.93\pm0.60$ & $-0.49\pm0.08$ & $5.56\pm0.41$  & $-0.46\pm0.10$ & $8.12\pm0.89$    \\
$<n_{b}^{b}>-n_{g}$       & $-0.21\pm0.05$ & $3.06\pm0.30$ & $-0.09\pm0.02$ & $2.67\pm0.13$ & $0.01\pm0.04$  & $3.03\pm0.20$ & $-0.27\pm0.04$ & $3.53\pm0.15$  & $-0.30\pm0.07$ & $3.24\pm0.36$    \\
$<n_{b}^{f}>-n_{g}$       & $-0.45\pm0.05$ & $6.17\pm0.19$ & $-0.21\pm0.04$ & $6.27\pm0.23$ & $-0.16\pm0.05$ & $6.81\pm0.29$ & $-0.48\pm0.05$ & $6.71\pm0.24$  & $-0.41\pm0.10$ & $5.86\pm0.43$  \\
$<n_{g}^{b}>-n_{h}$       & $0.13\pm0.10$  & $-0.62\pm0.94$& $0.10\pm0.01$  & $-0.18\pm0.13$& $0.02\pm0.01$  & $0.30\pm0.14$ & $-0.02\pm0.02$ & $1.18\pm0.26$  & $0.04\pm0.01$  & $-0.07\pm0.07$     \\
$<n_{g}^{f}>-n_{h}$       & $0.42\pm0.17$  & $-1.16\pm1.76$& $0.35\pm0.02$  & $-0.54\pm0.27$& $0.20\pm0.02$  & $0.77\pm0.31$ & $0.14\pm0.04$  & $0.81\pm0.39$  & $0.60\pm0.01$  & $-1.01\pm0.20$      \\
$<n_{b}^{b}>-n_{h}$       & $0.31\pm0.14$  & $-0.77\pm1.34$& $0.16\pm0.02$  & $0.31\pm0.20$ & $0.23\pm0.02$  & $-0.06\pm0.23$& $0.26\pm0.03$  & $-0.22\pm0.29$ & $0.08\pm0.01$  & $0.65\pm0.10$    \\
$<n_{b}^{f}>-n_{h}$       & $0.18\pm0.15$  & $2.13\pm1.64$ & $0.48\pm0.03$  & $-0.48\pm0.31$& $0.48\pm0.02$  & $-0.26\pm0.32$& $0.64\pm0.03$  & $-2.08\pm0.35$ & $0.27\pm0.01$  & $0.36\pm0.17$  \\
$<n_{b}^{f}>-n_{b}^{b}$   & $-0.14\pm0.12$ & $3.83\pm0.46$ & $0.33\pm0.08$  & $4.61\pm0.22$ & $0.09\pm0.01$  & $4.58\pm0.31$ & $0.37\pm0.11$  & $4.17\pm0.29$  & $0.82\pm0.08$  & $3.43\pm0.21$  \\
$<n_{g}^{b}>-n_{b}^{b}$   & $-0.17\pm0.09$ & $1.06\pm0.27$ & $-0.09\pm0.03$ & $1.20\pm0.10$ & $-0.02\pm0.03$ & $0.67\pm0.11$ & $-0.22\pm0.04$ & $1.64\pm0.13$  & $0.12\pm0.02$  & $0.48\pm0.06$       \\
$<n_{g}^{f}>-n_{b}^{b}$   & $-0.04\pm0.41$ & $0.62\pm1.45$ & $0.25\pm0.12$  & $-0.44\pm0.36$& $0.19\pm0.14$  & $-0.32\pm0.49$& $0.17\pm0.11$  & $-0.30\pm0.38$ & $0.05\pm0.30$  & $-0.02\pm0.71$         \\
$<n_{h}^{b}>-n_{b}^{b}$   & $0.83\pm0.09$  & $1.06\pm0.27$ & $0.90\pm0.03$  & $1.22\pm0.10$ & $0.98\pm0.03$  & $0.67\pm0.11$ & $0.77\pm0.04$  & $1.65\pm0.13$  & $1.12\pm0.02$  & $0.48\pm0.06$  \\
$<n_{h}^{f}>-n_{b}^{b}$   & $-0.57\pm0.20$ & $0.86\pm0.63$ & $0.12\pm0.10$  & $8.68\pm0.24$ & $0.52\pm0.11$  & $0.0002\pm0.31$& $-0.03\pm0.12$ & $7.60\pm0.30$  & $1.30\pm0.16$ & $11.444pm0.44$  \\
$<n_{b}^{b}>-n_{b}^{f}$   & $0.14\pm0.12$  & $1.84\pm0.56$ & $0.10\pm0.03$  & $1.74\pm0.15$ & $0.21\pm0.03$  & $1.83\pm0.19$ & $0.13\pm0.03$  & $1.93\pm0.19$  & $0.18\pm0.02$ & $1.12\pm0.10$  \\
$<n_{g}^{b}>-n_{b}^{f}$   & $-0.18\pm0.07$ & $1.40\pm0.33$ & $-0.04\pm0.02$ & $1.23\pm0.12$ & $-0.03\pm0.02$ & $0.77\pm0.12$ & $-0.13\pm0.02$ & $1.64\pm0.14$  & $0.07\pm0.01$ & $0.32\pm0.05$  \\
$<n_{g}^{f}>-n_{b}^{f}$   & $-0.87\pm0.10$ & $6.53\pm0.64$ & $-0.20\pm0.03$ & $4.64\pm0.19$ & $-0.05\pm0.04$ & $3.65\pm0.24$ & $-0.22\pm0.04$ & $3.50\pm0.22$  & $0.35\pm0.04$ & $6.83\pm0.20$ \\
$<n_{h}^{b}>-n_{b}^{f}$   & $0.12\pm0.09$  & $2.50\pm0.37$ & $0.06\pm0.03$  & $2.98\pm0.16$ & $0.20\pm0.02$  & $2.48\pm0.18$ & $0.05\pm0.03$  & $3.50\pm0.19$  & $0.23\pm0.02$ & $1.52\pm0.12$ \\
$<n_{h}^{f}>-n_{b}^{f}$   & $0.09\pm0.11$  & $6.71\pm0.65$ & $0.80\pm0.03$  & $4.63\pm0.20$ & $0.94\pm0.04$  & $3.68\pm0.25$ & $0.76\pm0.04$  & $3.58\pm0.22$  & $1.35\pm0.04$  & $6.82\pm0.21$ \\
$<n_{g}^{f}>-n_{g}^{b}$   & $1.31\pm0.36$  & $2.35\pm0.35$ & $0.42\pm0.07$  & $3.15\pm0.13$ & $0.54\pm0.11$  & $3.21\pm0.15$ & $0.32\pm0.10$  & $2.08\pm0.13$  & $1.68\pm0.25$  & $7.90\pm0.26$ \\
$<n_{b}^{b}>-n_{g}^{b}$   & $-0.85\pm0.18$ & $2.85\pm0.29$ & $-0.17\pm0.05$ & $2.45\pm0.10$ & $0.11\pm0.10$  & $3.06\pm0.14$ & $-0.44\pm0.07$ & $3.11\pm0.12$  & $0.33\pm0.07$  & $1.95\pm0.08$ \\
$<n_{b}^{f}>-n_{g}^{b}$   & $-0.75\pm0.20$ & $4.66\pm0.38$ & $-0.19\pm0.11$ & $5.56\pm0.17$ & $-0.30\pm0.21$ & $6.44\pm0.27$ & $-0.74\pm0.14$ & $5.90\pm0.21$  & $0.97\pm0.15$ & $4.57\pm0.18$ \\
$<n_{h}^{b}>-n_{g}^{b}$   & $0.09\pm0.18$  & $2.90\pm0.29$ & $0.83\pm0.05$  & $2.45\pm0.10$ & $1.11\pm0.10$  & $3.06\pm0.14$ & $0.56\pm0.07$  & $3.11\pm0.12$  & $1.33\pm0.08$  & $1.95\pm0.08$  \\
$<n_{h}^{f}>-n_{g}^{b}$   & $0.75\pm0.44$  & $6.99\pm0.31$ & $0.20\pm0.14$  & $8.74\pm0.19$ & $-0.05\pm0.27$ & $9.64\pm0.29$ & $-0.49\pm0.16$ & $8.02\pm0.23$  & $2.77\pm0.31$ & $12.41\pm0.36$ \\
$<n_{g}^{b}>-n_{g}^{f}$   & $0.20\pm0.05$  & $-0.05\pm0.15$& $0.09\pm0.02$  & $0.67\pm0.09$ & $0.03\pm0.03$  & $0.50\pm0.12$ & $0.12\pm0.03$  & $0.76\pm0.10$  & $0.06\pm0.01$  & $0.20\pm0.08$ \\
$<n_{b}^{b}>-n_{g}^{f}$   & $-0.17\pm0.08$ & $2.95\pm0.33$ & $-0.12\pm0.03$ & $2.68\pm0.13$ & $-0.02\pm0.06$ & $3.11\pm0.22$ & $-0.22\pm0.05$ & $3.13\pm0.14$  & $0.06\pm0.02$ & $1.67\pm0.13$ \\
$<n_{b}^{f}>-n_{g}^{f}$   & $-0.49\pm0.08$ & $5.76\pm0.30$ & $-0.23\pm0.06$ & $6.24\pm0.23$ & $-0.10\pm0.10$ & $6.58\pm0.33$ & $-0.25\pm0.06$ & $5.81\pm0.22$  & $0.26\pm0.03$ & $3.10\pm0.23$ \\
$<n_{h}^{b}>-n_{g}^{f}$   & $-0.06\pm0.08$ & $3.22\pm0.32$ & $-0.04\pm0.04$ & $3.41\pm0.15$ & $0.001\pm0.066$  & $3.65\pm0.23$ & $-0.12\pm0.05$ & $3.97\pm0.15$  & $0.12\pm0.02$ & $1.88\pm0.16$ \\
$<n_{h}^{f}>-n_{g}^{f}$   & $0.44\pm0.10$  & $5.91\pm0.32$ & $0.76\pm0.06$  & $6.26\pm0.23$ & $0.90\pm0.10$  & $6.59\pm0.34$ & $0.81\pm0.06$  & $5.70\pm0.22$  & $1.26\pm0.03$  & $3.10\pm0.23$ \\\hline
\end{tabular}
\end{small}
\end{sidewaystable}

\begin{sidewaystable}
Table 5. The fitting parameters of forward-backward multiplicity correlations for nucleus-emulsion interactions with $n_{h}\geq0$ using linear relation Eq.(1).
\begin{small}
\begin{tabular}{lcccccccccc}\hline
Correlation     & \multicolumn{2}{c}{150 A MeV $^{4}$He} & \multicolumn{2}{c}{290 A MeV $^{12}$C} & \multicolumn{2}{c}{400 A MeV $^{12}$C} & \multicolumn{2}{c}{400 A MeV $^{20}$Ne} & \multicolumn{2}{c}{500 A MeV $^{56}$Fe} \\
                          &      a           &    b            &       a          &      b          &       a          &        b        &       a       &       b        &       a       &       b  \\\hline
$<n_{g}^{b}>-n_{b}$       & $0.04\pm0.02$  & $0.24\pm0.05$ & $0.08\pm0.01$  & $0.14\pm0.02$ & $0.04\pm0.01$  & $0.03\pm0.01$ & $0.02\pm0.01$ & $0.42\pm0.04$  & $0.07\pm0.01$ & $0.05\pm0.01$    \\
$<n_{g}^{f}>-n_{b}$       & $-0.06\pm0.04$ & $1.63\pm0.13$ & $0.19\pm0.01$  & $0.99\pm0.04$ & $0.17\pm0.02$  & $1.01\pm0.07$ & $0.07\pm0.01$ & $0.88\pm0.06$  & $0.77\pm0.04$ & $1.65\pm0.12$    \\
$<n_{b}^{b}>-n_{g}$       & $-0.01\pm0.03$ & $0.84\pm0.08$ & $0.17\pm0.01$  & $0.58\pm0.03$ & $0.29\pm0.03$  & $0.86\pm0.06$ & $0.12\pm0.02$ & $1.01\pm0.04$  & $0.16\pm0.01$ & $0.44\pm0.04$        \\
$<n_{b}^{f}>-n_{g}$       & $-0.03\pm0.05$ & $1.74\pm0.12$ & $0.38\pm0.02$  & $1.58\pm0.06$ & $0.46\pm0.03$  & $1.94\pm0.11$ & $0.19\pm0.03$ & $1.88\pm0.07$  & $0.40\pm0.01$ & $0.86\pm0.06$          \\
$<n_{g}^{b}>-n_{h}$       & $0.08\pm0.01$  & $-0.07\pm0.04$& $0.08\pm0.00$  & $-0.05\pm0.01$& $0.04\pm0.00$  & $-0.07\pm0.02$& $0.09\pm0.01$ & $0.01\pm0.02$  & $0.04\pm0.00$  & $-0.02\pm0.01$     \\
$<n_{g}^{f}>-n_{h}$       & $0.31\pm0.02$  & $0.15\pm0.09$ & $0.30\pm0.01$  & $0.10\pm0.03$ & $0.24\pm0.01$  & $0.29\pm0.05$ & $0.23\pm0.01$ & $-0.07\pm0.03$ & $0.56\pm0.01$  & $-0.19\pm0.04$      \\
$<n_{b}^{b}>-n_{h}$       & $0.17\pm0.01$  & $0.04\pm0.06$  & $0.18\pm0.01$  & $-0.08\pm0.02$& $0.23\pm0.01$  & $-0.06\pm0.04$& $0.24\pm0.01$ & $0.03\pm0.04$  & $0.11\pm0.00$  & $0.12\pm0.02$    \\
$<n_{b}^{f}>-n_{h}$       & $0.33\pm0.02$  & $0.13\pm0.08$ & $0.45\pm0.01$  & $0.01\pm0.03$ & $0.47\pm0.01$ & $-0.01\pm0.05$ & $0.46\pm0.01$ & $-0.02\pm0.04$ & $0.28\pm0.01$  & $0.14\pm0.03$  \\
$<n_{b}^{f}>-n_{b}^{b}$   & $0.62\pm0.10$  & $1.19\pm0.09$ & $1.01\pm0.04$  & $1.46\pm0.05$ & $1.05\pm0.06$  & $1.45\pm0.09$ & $0.80\pm0.05$ & $1.32\pm0.06$  & $1.23\pm0.05$  & $1.27\pm0.07$  \\
$<n_{g}^{b}>-n_{b}^{b}$   & $-0.02\pm0.04$ & $0.37\pm0.05$ & $0.13\pm0.02$  & $0.26\pm0.02$ & $0.09\pm0.01$  & $0.10\pm0.02$ & $0.04\pm0.02$ & $0.46\pm0.03$  & $0.18\pm0.02$ & $0.15\pm0.02$       \\
$<n_{g}^{f}>-n_{b}^{b}$   & $0.06\pm0.12$  & $-0.02\pm0.16$& $0.09\pm0.05$  & $0.03\pm0.05$ & $0.09\pm0.07$  & $0.05\pm0.10$ & $0.06\pm0.05$ & $-0.04\pm0.08$ & $0.03\pm0.17$  & $-0.01\pm0.19$         \\
$<n_{h}^{b}>-n_{b}^{b}$   & $0.99\pm0.04$  & $0.37\pm0.05$ & $1.13\pm0.02$  & $0.26\pm0.02$ & $1.09\pm0.01$  & $0.10\pm0.02$ & $1.04\pm0.02$ & $0.46\pm0.03$  & $1.18\pm0.02$  & $0.15\pm0.02$  \\
$<n_{h}^{f}>-n_{b}^{b}$   & $0.66\pm0.12$  & $2.76\pm0.14$ & $1.46\pm0.06$  & $2.71\pm0.07$ & $1.47\pm0.08$  & $2.69\pm0.13$ & $0.95\pm0.07$ & $2.30\pm0.09$  & $2.79\pm0.12$ & $4.11\pm0.17$  \\
$<n_{b}^{b}>-n_{b}^{f}$   & $0.25\pm0.04$  & $0.45\pm0.07$ & $0.26\pm0.01$  & $0.29\pm0.02$ & $0.34\pm0.02$  & $0.43\pm0.05$ & $0.28\pm0.02$ & $0.59\pm0.04$  & $0.26\pm0.01$ & $0.41\pm0.03$  \\
$<n_{g}^{b}>-n_{b}^{f}$   & $0.04\pm0.02$  & $0.28\pm0.05$ & $0.09\pm0.01$  & $0.16\pm0.02$ & $0.06\pm0.01$  & $0.04\pm0.02$ & $0.03\pm0.01$ & $0.43\pm0.04$  & $0.09\pm0.01$ & $0.08\pm0.02$  \\
$<n_{g}^{f}>-n_{b}^{f}$   & $-0.18\pm0.04$ & $1.78\pm0.11$ & $0.19\pm0.01$  & $1.12\pm0.04$ & $0.24\pm0.02$  & $1.09\pm0.07$ & $0.10\pm0.02$ & $0.93\pm0.06$  & $0.81\pm0.04$ & $2.10\pm0.10$ \\
$<n_{h}^{b}>-n_{b}^{f}$   & $0.29\pm0.04$  & $0.73\pm0.08$ & $0.35\pm0.01$  & $0.46\pm0.03$ & $0.38\pm0.01$  & $0.50\pm0.05$ & $0.32\pm0.02$ & $1.01\pm0.05$  & $0.33\pm0.01$ & $0.51\pm0.03$ \\
$<n_{h}^{f}>-n_{b}^{f}$   & $0.82\pm0.04$  & $1.79\pm0.11$ & $1.20\pm0.02$  & $1.11\pm0.04$ & $1.24\pm0.02$  & $1.08\pm0.07$ & $1.10\pm0.02$ & $0.93\pm0.06$  & $1.83\pm0.04$  & $2.08\pm0.10$ \\
$<n_{g}^{f}>-n_{g}^{b}$   & $0.55\pm0.17$  & $1.36\pm0.09$ & $0.82\pm0.05$  & $1.36\pm0.04$ & $1.01\pm0.09$  & $1.61\pm0.07$ & $0.62\pm0.06$ & $0.85\pm0.04$  & $3.30\pm0.22$  & $3.60\pm0.13$ \\
$<n_{b}^{b}>-n_{g}^{b}$   & $-0.08\pm0.06$ & $0.89\pm0.07$ & $0.30\pm0.04$  & $0.81\pm0.03$ & $0.72\pm0.08$  & $1.33\pm0.06$ & $0.08\pm0.04$ & $1.21\pm0.05$  & $0.71\pm0.06$ & $0.95\pm0.04$ \\
$<n_{b}^{f}>-n_{g}^{b}$   & $0.15\pm0.11$  & $1.65\pm0.10$ & $0.94\pm0.08$  & $2.05\pm0.06$ & $1.35\pm0.17$  & $2.75\pm0.11$ & $0.29\pm0.09$ & $2.19\pm0.08$  & $1.89\pm0.12$ & $2.12\pm0.08$ \\
$<n_{h}^{b}>-n_{g}^{b}$   & $0.92\pm0.07$  & $0.89\pm0.07$ & $1.30\pm0.04$  & $0.81\pm0.03$ & $1.72\pm0.08$  & $1.33\pm0.06$ & $1.08\pm0.04$ & $1.21\pm0.05$  & $1.71\pm0.06$  & $0.95\pm0.04$  \\
$<n_{h}^{f}>-n_{g}^{b}$   & $0.99\pm0.25$  & $3.00\pm0.12$ & $1.76\pm0.11$  & $3.40\pm0.08$ & $2.28\pm0.22$  & $4.37\pm0.15$ & $0.91\pm0.11$ & $3.04\pm0.10$  & $5.24\pm0.27$ & $5.72\pm0.19$ \\
$<n_{g}^{b}>-n_{g}^{f}$   & $0.05\pm0.02$  & $0.23\pm0.04$ & $0.13\pm0.01$  & $0.16\pm0.02$ & $0.09\pm0.01$  & $0.09\pm0.02$ & $0.20\pm0.02$ & $0.27\pm0.03$  & $0.07\pm0.01$  & $0.05\pm0.02$ \\
$<n_{b}^{b}>-n_{g}^{f}$   & $0.08\pm0.04$  & $0.77\pm0.08$ & $0.20\pm0.02$  & $0.60\pm0.03$ & $0.31\pm0.03$  & $0.94\pm0.07$ & $0.17\pm0.03$ & $1.04\pm0.05$  & $0.15\pm0.01$  & $0.51\pm0.04$ \\
$<n_{b}^{f}>-n_{g}^{f}$   & $0.01\pm0.05$  & $1.73\pm0.11$ & $0.45\pm0.03$  & $1.66\pm0.06$ & $0.63\pm0.06$  & $1.91\pm0.12$ & $0.33\pm0.04$ & $1.91\pm0.07$  & $0.41\pm0.02$ & $0.90\pm0.06$ \\
$<n_{h}^{b}>-n_{g}^{f}$   & $0.13\pm0.05$  & $1.00\pm0.09$ & $0.34\pm0.02$  & $0.75\pm0.04$ & $0.40\pm0.04$  & $1.03\pm0.08$ & $0.34\pm0.03$ & $1.34\pm0.05$  & $0.22\pm0.01$ & $0.55\pm0.04$ \\
$<n_{h}^{f}>-n_{g}^{f}$   & $1.01\pm0.06$  & $1.72\pm0.12$ & $1.45\pm0.03$  & $1.66\pm0.06$ & $1.65\pm0.06$  & $1.89\pm0.12$ & $1.38\pm0.04$ & $1.88\pm0.07$  & $1.41\pm0.02$ & $0.90\pm0.06$ \\\hline
\end{tabular}
\end{small}
\end{sidewaystable}

\section{Summary}
The forward-backward multiplicity and correlations of black track particles and grey track particles emitted in 150 A MeV $^{4}$He-emulsion, 290 A MeV $^{12}$C-emulsion, 400 A MeV $^{12}$C-emulsion, 400 A MeV $^{20}$Ne-emulsion and 500 A MeV $^{56}$Fe-emulsion interactions are investigated. It is found that the forward and backward averaged multiplicity of grey, black and heavily ionized track particle increase with the increase of target size. Averaged multiplicity of forward black track particle, backward black track particle, and backward grey track particle do not depend on the projectile size and energy, but the averaged multiplicity of forward grey track particle increases with increase of projectile size and energy. The backward grey track particle multiplicity distribution follows an exponential decay law and the decay constant decreases with increase of target size. The general characteristics of backward-forward multiplicity correlations of black, grey and heavily ionized track particles are discussed, which can be well explained by the geometrical picture and the cascade evaporation model of high nucleus-nucleus collisions.

\section{Acknowledgements}
This work has been supported by the Chinese National Science Foundation under Grant No: 11075100 and the Natural Foundation of Shanxi Province under Grant 2011011001-2, the Shanxi Provincial Foundation for Returned Overseas Chinese Scholars, China (Grant No. 2011-058). We gratefully acknowledge the staffs of the HIMAC for providing the beam to expose the stacks.

\end{document}